\documentclass[10pt,conference]{IEEEtran}
\usepackage{cite}
\usepackage{amsmath,amssymb,amsfonts}
\usepackage{algorithmic}
\usepackage{graphicx,epstopdf}
\usepackage{textcomp}
\usepackage{xcolor}
\def\BibTeX{{\rm B\kern-.05em{\sc i\kern-.025em b}\kern-.08em
    T\kern-.1667em\lower.7ex\hbox{E}\kern-.125emX}}

\usepackage{url}
\usepackage{soul}
\usepackage{booktabs}
\DeclareMathOperator*{\argmax}{arg\,max}
\DeclareMathOperator*{\argmin}{arg\,min}

\begin{document}

\onecolumn

{\LARGE IEEE Copyright Notice} \\

\copyright 2019 IEEE. Personal use of this material is permitted. Permission from IEEE must be obtained for all other uses, in any current or future media, including reprinting/republishing this material for advertising or promotional purposes, creating
new collective works, for resale or redistribution to servers or lists, or reuse of any copyrighted component of this work in other works. \\

{\large Accepted to be Published in: Proceedings of the IEEE Computer Society Annual Symposium on VLSI (ISVLSI), July 15-17, 2019, Miami, Florida, U.S.A.}

\twocolumn

\title{Machine Learning based IoT Edge Node Security Attack and Countermeasures}

\author{
\IEEEauthorblockN{Vishalini R. Laguduva, Sheikh Ariful Islam, Sathyanarayanan Aakur, Srinivas Katkoori and Robert Karam}
\IEEEauthorblockA{Department of Computer Science and Engineering \\
University of South Florida\\
Tampa, FL 33620 \\
{\{vishalini, sheikhariful, saakur, katkoori, rkaram\}@mail.usf.edu}}
}

\maketitle

\begin{abstract}
Advances in technology have enabled tremendous progress in the development of a highly connected ecosystem of ubiquitous computing devices collectively called the Internet of Things (IoT). Ensuring the security of IoT devices is a high priority due to the sensitive nature of the collected data. 
Physically Unclonable Functions (PUFs) have emerged as critical hardware primitive for ensuring the security of IoT nodes. 
Malicious modeling of PUF architectures has proven to be difficult due to the inherently stochastic nature of PUF architectures. 
Extant approaches to malicious PUF modeling assume that {\em a priori} knowledge and physical access to the PUF architecture is available for malicious attack on the IoT node. 
However, many IoT networks make the underlying assumption that the PUF architecture is sufficiently tamper-proof, both physically and mathematically. In this work, we show that knowledge of the underlying PUF structure is not necessary to clone a PUF. We present a novel non-invasive, architecture independent, machine learning attack for strong PUF designs with a cloning accuracy of 93.5\% and improvements of up to 48.31\% over an alternative, two-stage brute force attack model. 
We also propose a machine-learning based countermeasure, \textit{discriminator}, which can distinguish cloned PUF devices and authentic PUFs with an average accuracy of 96.01\%.  The proposed discriminator can be used for rapidly authenticating millions of IoT nodes remotely from the cloud server.
\end{abstract}

% \begin{IEEEkeywords}
% IoT, Internet of Things, Security, PUFs, Machine Learning
% \end{IEEEkeywords}

\section{Introduction and Motivation}

Evolution of technology has resulted in a highly connected ecosystem of ubiquitous computing devices that work together seamlessly to collect, process and analyze large amounts of data to aid in human-centric decision making. 
Collectively called the Internet of Things (IoT), the collection of wearable devices, sensors and embedded systems (to name a few) have enabled automated decision making for improving quality of life. 
% Given the extent to which such devices are integrated into our daily lives, adversarial attacks on such devices can lead to high levels of security and trust issues. 
Given the highly integrated nature of IoT devices, adversarial attacks can lead to high levels of security and trust issues. 
Ensuring the security of IoT devices is a high priority due to the sensitive nature of the collected data\cite{cam2016can,ray2016security}. 
However, this comes with it a set of challenges: (1) IoT devices are typically resource-constrained, thus requiring high energy efficient security protocols; (2) their highly distributed nature can provide easy physical access to the node and (3) the highly connected nature of IoT framework requires fast and secure security protocols. 

Traditional approaches to cryptography, while effective, have not proven to be sufficiently lightweight and fast for IoT device authentication. 
Thus, hardware-based security protocols have emerged as viable alternatives 
%to provide lightweight and highly secure authentication protocols 
for IoT device registration and authentication. 
%Typical implementation of such approaches is to store the key in silicon-based memory devices, which can be prone to physical attacks. 
%There have been several approaches to prevent such attacks through hardware primitives such as obfuscation.
%i.e. hiding data in complex chip layouts or below dense metal structures for visual obscurity. 
% 
Recent efforts have shifted to leveraging the inherent randomness induced in silicon devices during the manufacturing process as the secret key, opposed to the traditional binary key stored in silicon devices, which can be susceptible to physical attacks. 
Such approaches, called Physically Unclonable Functions (PUFs), have helped provide a higher level of security against direct physical attacks. This alleviates the need for costly physical protection measures.
PUFs have become increasingly popular and have been used for IoT device authentication~\cite{chatterjee2017puf, chatterjee2018building, aman2018token, aman2019hardware, braeken2018puf} and other security tasks \cite{Pappu2026, Ruhrmairmulti}.
% as well as in a variety of hardware-based security tasks such as for identification and authentication \cite{Pappu2026}, digital rights management  \cite{Gassend:2002}, bit-commitment protocol \cite{Pappu2026}, and secure multi-party communication \cite{Ruhrmairmulti}. 

Silicon-based PUF devices \cite{Pappu2026} are easily fabricated, physical structures that leverage the stochastic nature of the manufacturing process to create physically unclonable, unique identifiers for each manufactured unit. 
This typically results in a one-way function. Given an electronic stimulus, the response of a PUF device is an unpredictable, repeatable function. This response identifies each device with a unique signature. 
% This is largely attributed to the interaction of the external stimulus and the physical structure of the PUF. 
This interaction is termed as the Challenge-Response Pair (CRP), where the challenge is the external stimulus and the PUF's reaction is termed as the response. 
%The unpredictable nature of the PUF, which can be highly sensitive to noise, error correction circuits \cite{maes2009low} are used to reduce the uncertainty in the PUF's response to make it more reliable. 
% A PUF with sufficiently large challenge response pairs are called strong PUFs and are typically chosen for most practical security applications. 

\begin{figure}
\vspace{-2ex}
%\centering
\includegraphics[width=0.95\columnwidth]{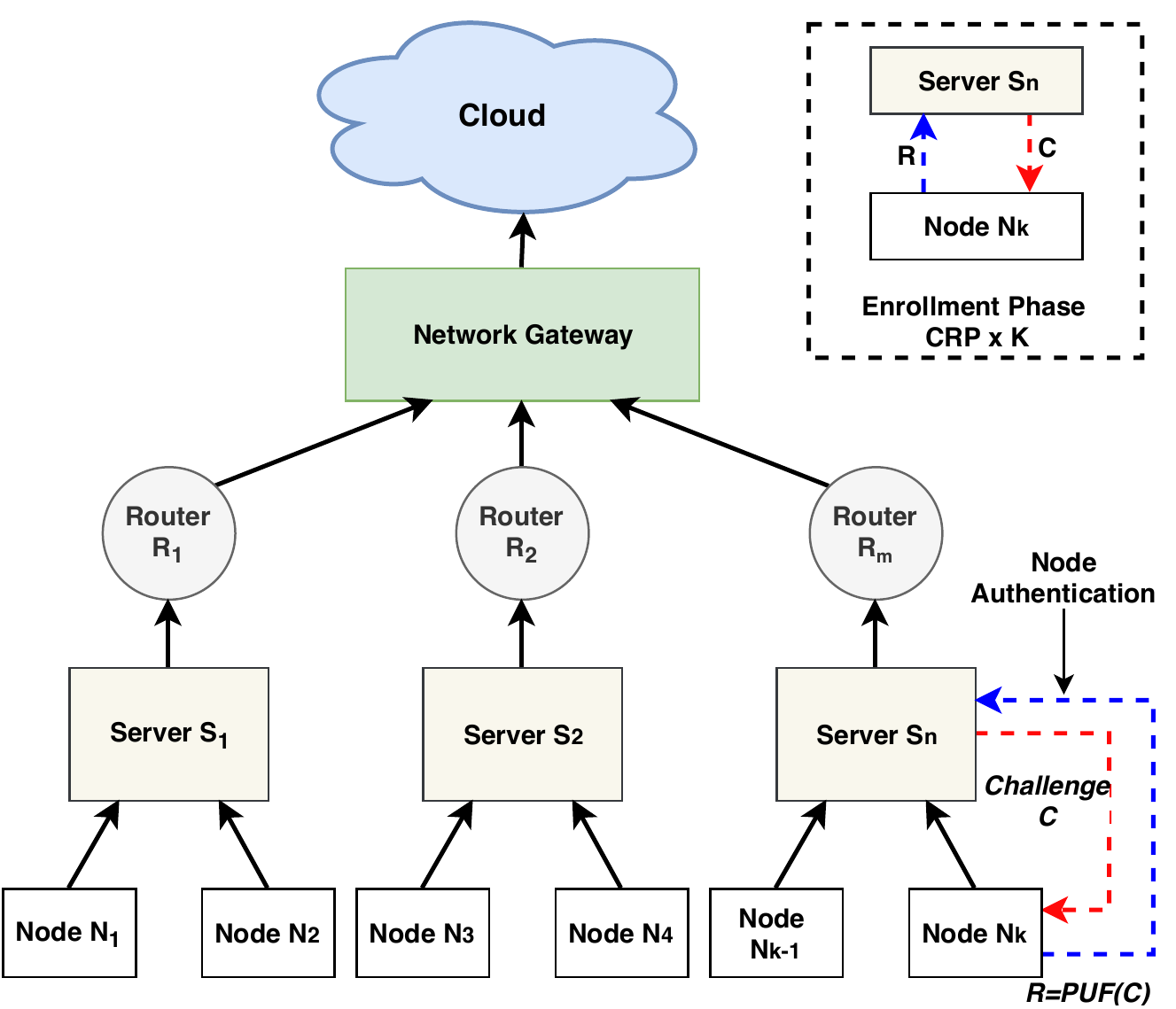}
\caption{Typical IoT architecture is illustrated here. The dashed lines indicate IoT device authentication protocols. IoT device enrollment is done by populating and authenticating CRPs ``$K$'' times.} 
\label{fig:overallArch}
\vspace{-1ex}
\end{figure}

Using PUFs for IoT security protocols typically involves an initial enrollment phase and an authentication protocol during the actual data exchange. 
Figure \ref{fig:overallArch} illustrates the typical architecture of an IoT network and the generic enrollment protocol. 
A typical IoT network consists of remote, resource-constrained data nodes ($N_1, N_2, N_3 \ldots N_k$) connected to static server nodes ($S_1, S_2, S_3 \ldots S_n$) that transfer the acquired data to the cloud using routers ($R_1, R_2, R_3 \ldots R_m$). 
The data is transmitted from the routers to the cloud using a network gateway. 
IoT edge nodes can range from simple sensors to complex systems with processor, memory, communication etc. Strong PUFs implemented in complex IoT nodes are subject to attacks, which is the focus of this work. When a data node is added to the IoT network, an enrollment phase is executed to create a CRP database for the PUF within the data node. 
This database of CRPs is used in the authentication phase when two nodes corresponding to the same server node want to communicate. 
The common server node authenticates both data nodes, generates security key pairs and helps secure key sharing.

\textbf{Security Assumptions:} Following the protocols established in \cite{ostrovsky2013universally}, extant IoT networks using PUF authentication~\cite{chatterjee2017puf, chatterjee2018building, aman2018token, aman2019hardware, braeken2018puf} make the following underlying assumptions: (1) cloning a PUF architecture, either physically or mathematically is a difficult problem,
% , especially if the underlying architecture is unknown
(2) an adversary has unrestricted physical access to the communication channel, (3) the challenge-response characteristics of the PUF within the data IoT node is an implicit property and is not accessible to an adversary and (4) the attacker can obtain access to the database of CRPs through malicious software attacks, though explicit knowledge of the secret keys.
Given these security assumptions, the goal of the adversary becomes straightforward: it must be able to spoof the server nodes into accepting a malicious node on behalf of the original data nodes without actual possession of the node in question. 
Any physical intrusions can compromise the integrity of the PUF and hence render the attack futile. 
The underlying stochastic nature of PUFs and the above constraints lend itself to a solid security protocol that can be hard to breach. 
However, advances in machine learning have led to a vast majority of the non-invasive attacks explored. 
Machine learning-based approaches can be characterized by the application of a learned mathematical model on a collected subset of valid CRPs. The curation of such data is typically assumed to be an eavesdropping protocol.
%, which is not an unreasonable assumption. % Hence, CRPs, unique to a PUF architecture, bring both challenges and opportunities. 
Prior works, especially the pioneering work of R\"{u}hmair et al\cite{Ruhrmair2010}, have shown great success in cloning PUFs, gaining cloning accuracy of up to 99.99\%. 
Such success does come with a caveat - the underlying architecture must be known {\textit{a priori}}, either through invasive physical intrusions or explicit architecture knowledge. 

Today's IoT nodes are designed such that they are tamper-proof \cite{ishai2006private} \cite{yang2015protecting} which makes it difficult or impossible for micro-probing. 
Even if the attacker is successful in micro-probing, given the myriad of PUF architectures in literature, extracting information on the underlying PUF architecture is extremely difficult. 
Hence, earlier ML-based PUF attacks with the assumption of knowing underlying architecture are either not practical or extremely difficult to stage. 
In this work, we present, for the first time, an ML-based attack that does not require PUF architecture information.  We also present a countermeasure for this attack that can be effectively used to remotely evaluate an IoT node's trust level. 

%Given the above, 

To overcome such limitations, we focus on an architecture independent attack, that assumes no prior knowledge of the PUF architecture in the system. 
We show that observed CRPs are sufficient to improve cloning accuracy of a strong PUF irrespective of the underlying architecture.
% while PUF implements the authentication protocol. 
The attack can simulate PUF network without knowing underlying PUF architecture. 
To evaluate the effectiveness of our approach, we compare against a brute force attack model (Section \ref{sec:BruteForce}) that leverages the current advances in PUF-architecture cloning. We leverage architecture-specific-cloning \cite{Ruhrmair2010} through a cascaded framework of (1) PUF architecture identification, (2) employ architecture specific cloning models, and (3) evaluate the prediction accuracy of the model by combining the architecture classification accuracy and the cloning accuracy in a harmonic mean.
% To employ architecture-specific cloning, such as those proposed by \cite{Ruhrmair2010}, we first need to identify the PUF architecture. 
% We introduce a mathematical model to use the PUF's response to identify its underlying architecture and then employ the cloning models to attack the PUF.
% We compute the prediction accuracy of the model by combining the architecture classification accuracy and the cloning accuracy in a harmonic mean. 
% The approach is explained in detail in Section \ref{sec:BruteForce}. 
% However, in a \textit{controlled PUF}\cite{Gassend:2002}, both machine learning-based attack models cannot be a viable approach as the challenge-response of the underlying PUF is not directly accessible, but is rather protected by a controlling logic interface. 
% All challenges are pre-processed by the controlling logic before they are input to the PUF and all responses of the PUF is post processed, adding a level of complexity to the security of the PUF.

Inspired from the pioneering work of Goodfellow \emph{et al} \cite{goodfellow2014generative} on Generative Adversarial Networks (GANs), we propose a machine learning-based defense, a \textit{discriminator}, to identify the possibility of cloning using any ML-based attack non-invasive attack. 
Extant countermeasures\cite{8031539, 6714458} to ML-based cloning have focused on creating complex, cloning resistant PUF architecture. 
% Prior approaches to ML-based cloning countermeasures provided added security through obfuscation \cite{8031539}
% % \cite{8031539,7457162}, 
% % meta-algorithms based machine learning \cite{7495550}, 
% and sub-string matching \cite{6714458}, to name a few. 
% However, these countermeasure methods are based on creating complex PUF architectures. 
As we enter into a more realizable IoT ecosystem, complex PUF architectures may not be suitable for lightweight IoT systems.
Hence, we propose a lightweight, probabilistic identification of cloning through machine learning. To the best of the authors' knowledge, this is the first such framework for the non-invasive attack of PUF-based IoT network authentication schemes and a proposed mechanism to differentiate original PUFs from cloned ones. In short, our paper makes the following novel contributions:
\begin{itemize}
\item propose a non-invasive, architecture independent cloning attack on string PUFs,
\item show that a brute force attack on strong PUFs to identify the PUF architecture for cloning is increasingly complex and hence not trivial for feasible cloning, and
\item propose a probabilistic, discriminator model to bolster the security of the CRP protocol by identifying possible instances of cloning attacks.
\end{itemize}

% As such, the techniques discussed in this paper can be applied to a wide variety of PUFs. 

The rest of this paper is organized as follows. We briefly review extant machine learning attacks on PUFs and corresponding countermeasures in Section \ref{sec:ref}. We describe and evaluate a baseline, brute-force approach in Section \ref{sec:BruteForce}, followed by a description of the proposed attack and discriminator approach in Section \ref{sec:approach}. We present our empirical evaluation of the proposed approach in Section \ref{sec:results} and conclude with a discussion on the proposed approach in Section \ref{sec:conclusions}.
% Section \ref{sec:ref} briefly reviews related works of machine learning attack and defense mechanisms of strong PUFs. 
% A brute force attack is introduced and evaluated in Section \ref{sec:BruteForce}. 
% The proposed attack method and discriminator model are introduced in Section \ref{sec:approach}, followed by experimental evaluation of the proposed approach in Section \ref{sec:results}. 
% Finally, we conclude with discussion on the feasibility of the proposed approach in Section \ref{sec:conclusions}.
\section{Background and Related Work}
\label{sec:ref}

% %Table1
% \begin{table*}[t]
% \caption{Comparison of existing ML attacks (best results through simulation) and Modeling Resistant technique on strong PUFs}
% % \vspace{-3ex}
% \label{tab:comparison}
% \begin{center}
% \resizebox{0.9\textwidth}{!}
% {\begin{tabular}{|c|c|c|c|c|c|c|}
% \hline
% Strong PUFs & ML Attack Algorithms & ML resistance technique  \\
% \hline
% APUF  & LR \cite{Ruhrmair2010}, SVM \cite{6412622}, ANN \cite{6412622},  & Challenges Randomization \cite{7835567}, Multiple Input  \\
%   &  Reverse Fuzzy Extractor \cite{7096998} & Signature Register \cite{8050671}, Response Obfuscation \cite{8031539}  \\
% \hline
% Feed-Forward APUF & ES \cite{Ruhrmair2010}, Multi-layer Perceptrons \cite{8073845} & Gradient Boosting \cite{7495550} \\
% \hline
% XOR APUF &  LR \cite{Ruhrmair2010} & Function Composition \cite{7495550}, Response Obfuscation \cite{8031539}   \\
% \hline
% Lightweight PUF & LR \cite{Ruhrmair2010}  & --\\
% \hline
% Ring Oscillator PUF & --  & -- \\
% \hline
% Current-based PUF & ES \cite{6974725}  &  SVM \cite{7092470}\\
% \hline
% \end{tabular}}
% \end{center}
% % \vspace{-3ex}
% \end{table*}

In this section, we briefly summarize extant work on machine-learning based attack and prevention techniques in the strong PUF design.
% Table \ref{tab:comparison} gives an overview of the existing ML attacks and modeling PUFs that are resistant to the ML attacks.

\textbf{Strong PUFs:} 
A strong PUF can support a large number of complex CRPs with physical access to the PUF for a query such that an attacker cannot generate correct response given finite resources and time \cite{6823677,6800561,8357321}. While a weak PUF has only few CRPs which makes it difficult for the attack and prediction techniques, hence in this paper we consider strong PUF. The number of CRPs of strong PUFs can grow exponentially depending on the number of module blocks available for generating responses for a large number of corresponding challenges. Error due to noise in the response of PUF can be minimized using helper data \cite{helper,8607170}. 
% The inherent CRPs of strong PUFs can grow exponentially depending on the number of module blocks available for response generation with large possibilities of corresponding challenges. Error induced by noise on the response of PUF can be minimized using helper data \cite{helper,8607170}. 
% For completeness, we assume such error-correction mechanism incorporating temperature, voltage, and aging variations are already present in the PUF to be cloned. A strong PUF does not contain read-out protection scheme assuming an attacker has to enumerate a large number of CRPs. 
% Hence, it makes an invasive attack infeasible while impelling attacker to apply ML-based techniques to be successful beyond the underlying complexity of strong PUFs. 
For a detailed analysis of constructions and description of strong PUFs, we refer the reader to \cite{6823677}.

\textbf{ML Attacks on PUFs:} 
R\"{u}hrmair \emph{et al.} \cite{Ruhrmair2010} proposed an ML-based attack on strong PUFs based on a predictive model. 
The authors were able to clone the functionality of the underlying PUF given the PUF model by evaluating model parameters using LR with RProp and ES. 
% The cloning model of the target PUF was trained for coefficients estimation (training set ranges from 10000 to 500000) and tested on the CRPs for the prediction rate. 
Though the method was quite successful in cloning, the attacker needs to know the underlying PUF architecture and the corresponding signature function.
While it is reasonable to assume that CRPs can be obtained by eavesdropping or other interfaces \cite{6800561}, it is not always possible to ascertain the underlying PUF model without physical access to the PUF. 
% R\"{u}hrmair \emph{et al.} \cite{6587277} validated the numerically simulated CRPs with the CRPs generated with silicon data (FPGA and ASIC). 
% Although the presented attacks work better under a given PUF size and architectural complexity, an attacker should have the idea of underlying PUF architecture to make the generated clone samples match the statistics of the real CRPs. 
There have also been other approaches such as PAC\cite{Ganji} and hybrid methods\cite{cryptoeprint:2013:851} that have successfully cloned PUFs using a combination of ML and invasive techniques.
% Recently, the combined ML and side-channel (timing and power) present an improved hybrid attack surface \cite{cryptoeprint:2013:632,cryptoeprint:2013:851}. 
% A mathematical model-free ML attack using PAC (Probably Approximately Correct) learning framework has been proposed in \cite{Ganji}. The authors presented that an influential bit, if present in stable PUF response, can predict the future response corresponding to a challenge with low probability. 
% Although the framework achieves a good score, the framework cannot be widely applied to strong PUFs that have been tackled in this paper.

\textbf{ML resistant PUFs:} 
The linear additive behavior of Arbiter PUF (APUF) has made it an ideal target for ML attack. 
Hence, higher non-linearity in a given PUF architecture can improve the uniqueness and randomness with increased defense against modeling attack.
Other approaches to ML resistant PUFs have been randomized challenges\cite{7835567}, obfuscation\cite{8031539}, sub-string based challenges\cite{6714458}.
% Randomized challenges \cite{7835567} to PUF and obfuscating PUF responses \cite{8031539} have also been proposed. 
% Rostami \emph{et al.} presented a prover-verifier framework for successful authentication based on a subset of response substring \cite{6714458}. 
% Vijayakumar \emph{et al.} proposed to utilize bagging and boosting ML algorithms to improve the accuracy of classifier given sufficient entropy of cascading PUFs \cite{7495550}.

% The majority of works describing ML resistant PUFs employ clearly defined architecture and adequately large CRPs for the training process. The randomness and uniqueness, instead, deteriorate substantially when CRPs that do not belong to original CRPs for a particular PUF are used as the case we are tackling in this work. Indeed, we present a discriminator model that do permit the investigation of CRPs received at a PUF challenge-response interface to lower the attacker attempt in reverse engineering the PUF model.

% 2.a. What problem are the researchers in this paper
% trying to solve?
% 2.b. What is the scope of their work?
% 2.c. What impactful and innovative ideas do the
% authors show?

\section{Brute Force Attack on Strong PUFs}\label{sec:BruteForce}
% In this section, we describe the brute force attack that we formulate for evaluating the effectiveness of the proposed solution. 
% We begin with the motivation for a brute force attack and basic assumptions that are required for applying the cloning models. 
% We then continue with discussion on the mathematical models employed for identifying the underlying PUF architecture and the subsequent results. 
% We conclude with advantages and limitations of the proposed approach. 

% \subsection{Motivation and Basic Assumptions}
The success of the proposed models by R\"{u}hmair et al \cite{Ruhrmair2010} allow us to successfully clone strong PUF models with a prediction accuracy of 99.9\%. 
However, to use it in a non-invasive manner, we would first need to identify the underlying PUF architecture, as the approaches in \cite{Ruhrmair2010} require intimate knowledge of the PUF architecture such as PUF type, number of stages and number of XOR gates, to name a few. 
\begin{figure}
\centering
\includegraphics[width=0.85\columnwidth]{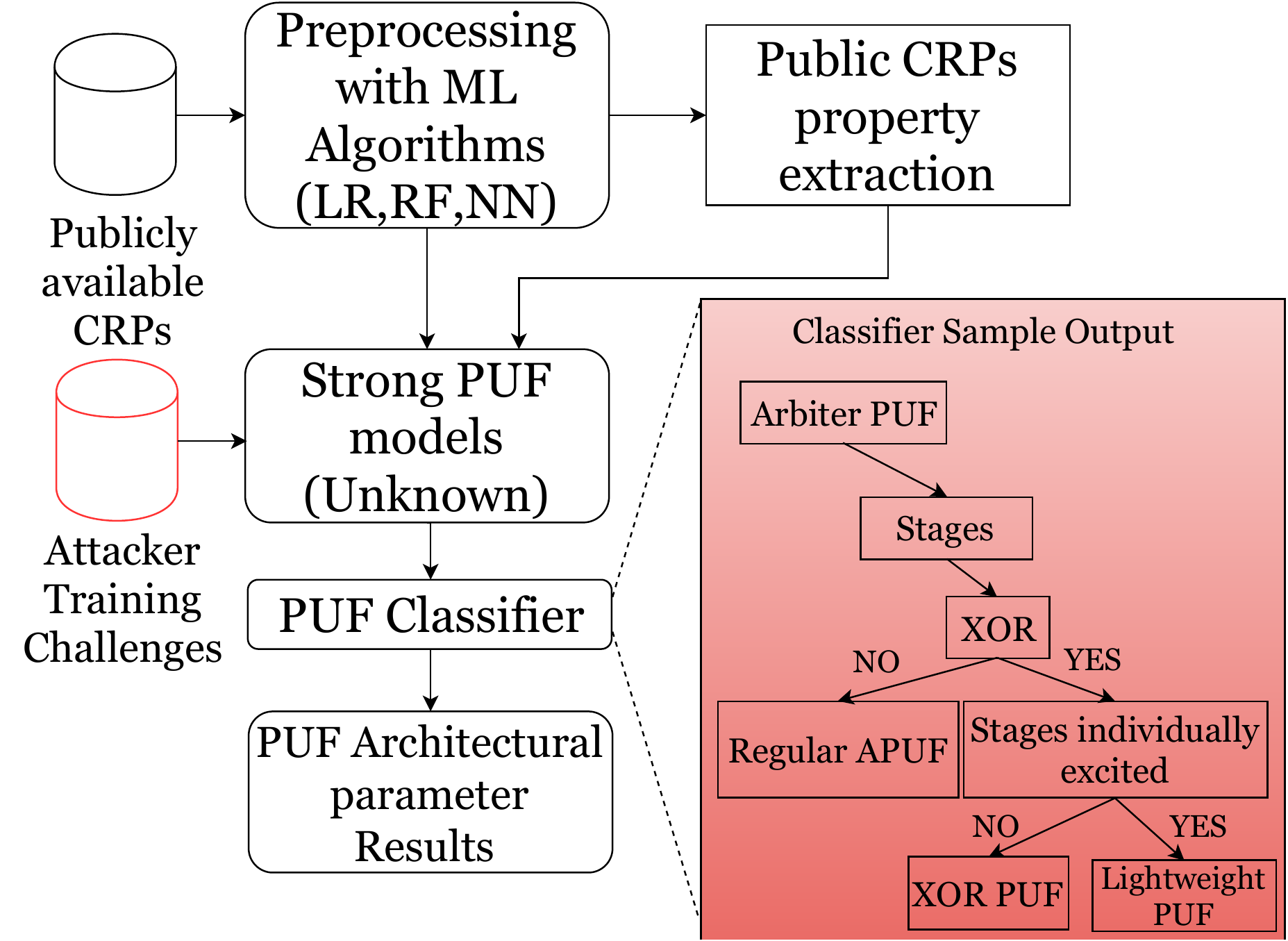}
\caption{The proposed attack model on oblivious PUFs architecture. The brute force attack has an additional PUF architecture detection process as indicated by the block in red.}
\label{fig:attack}
\vspace*{-3ex}
\end{figure}

To address this, we propose the use of a machine learning model to identify the PUF architecture through observation of the challenge-response pairs, as illustrated in Figure \ref{fig:attack}.
One major assumption in this approach is that there exists a subset of challenges $\Tilde{C} \in C$ that is valid for all PUF architectures in a given network, where $C$ is the collection of all valid CRPs.
Given the number of PUF architectures and their use for authentication, this is not an unreasonable assumption. 

\subsection{Identifying PUF architectures}
Given the set of challenges $\Tilde{C}$, we can observe the set of valid responses $R_{c_i}$ for each PUF architecture $c_i \in C_{puf}$, where $C_{puf}$ is the set of all known PUF architectures described in Section \ref{sec:ref}. Hence, the objective of the classification is to learn a function $f_c$ which maximizes the probability 
\begin{equation}
    \argmax_{\Tilde{C}_i\in\Tilde{C}} P(c_i | \Tilde{C}_i, R_{c_i})
    \label{eqn:classEqn}
\end{equation}
where the objective is the find the PUF architecture $c_i$ given the challenge $\Tilde{C}_i$, and the subsequent response $R_{c_i}$. 
We use the following machine learning models as the basis for the function $f_c(\cdot)$: logistic regression, artificial neural network, and random forests. 
\subsection{Empirical Evaluation}
\begin{table}
\begin{center}
\caption{Brute Force Attack: PUF architecture classification performance and subsequent cloning accuracy.  }
\vspace*{-3ex}
% When compared to this approach, our approach achieves 93.5\%.}
\label{tab:classPUF}
\resizebox{\columnwidth}{!}{
\begin{tabular}{l|c|c}
\toprule
PUF Model  &   {PUF } &   {Cloning Rate (\%)}   \\
 & Classification Rate (\%) & \\\midrule\midrule
APUF           & 81.49\%  & 77.42\% \\ 
3 XOR APUF     & 76.53\%  & 72.71\% \\
4 XOR APUF     & 65.01\%  &  61.76\% \\     
5 XOR APUF     & 63.57\%  &  60.39\% \\
6 XOR APUF     & 61.31\%  &  58.25\% \\
LW 3 XOR APUF  &  76.91\% &  73.05\% \\
LW 4 XOR APUF  &  65.37\% &  62.10\% \\
LW 5 XOR APUF  &  59.32\% &  56.33\% \\

\bottomrule
\end{tabular}}
\end{center}
\vspace*{-6ex}
\end{table}
We evaluate the performance of the proposed brute force attack to identify the architecture of eight ($8$) common strong PUF architectures. 
We use a fixed number of randomly sampled $100$ CRPs for evaluation for each PUF architecture for a total of $800$ CRPs. We report average results from 5 different runs, with the test set sampled randomly each time. We curate a set of $100,000$ CRPs for training the classification model. 

As can be seen from Table \ref{tab:classPUF}, identifying the PUF architecture from an observed set of CRPs is not a trivial task. 
Even with $100\%$ cloning accuracy for a given PUF architecture, identifying the said architecture requires a large set of CRPs for training a model. 
The maximum performance that we were able to obtain was using the logistic regression model, which took 100 iterations to converge resulting in the maximum classification rate for Arbiter PUF architecture. 
There was a large confusion among different design variations of each PUF type. The prediction rate for XOR PUFs decreased as the complexity of the architecture increased.

% \subsection{Discussion}
% It can be seen that identifying the PUF architecture requires significant training resources - $100,000$ CRPs, while identifying the arbiter PUF with an average accuracy of $81.49\%$. The classifier performed worst on the lightweight PUFs, yielding a maximum identification accuracy for the 3 bit XOR lightweight PUF. 
% The identification rate also affected the cloning prediction rate of the brute force approach, as each mis-classified PUF architecture affecting the cloning quality. 
While the average cloning accuracy can be as high as $77.42\%$ (for the Arbiter PUF), the numbers can be misleading in practice. 
The performance of the two-stage attack model is rather low, considering the practical gap between the intra and inter Hamming Distance of PUF CRPs, this prediction rate cannot be considered to be successful cloning.% of the PUF architecture.
\section{Architecture Independent PUF Modeling}
\label{sec:approach}

In this section, we describe our proposed approach for a PUF-independent attack model on various PUF architectures by exploiting the CRP authentication protocol. 
We begin with a discussion on the use of machine learning models to capture the underlying correlation between challenge-response pairs to model the randomness unique to a given PUF architecture. 
We then follow with a discussion on defending against such attacks using complementary machine learning models.

\subsection{Attack Model}
\label{attack}
Each PUF is made unique through a digital signature characterized by its response to a given challenge. 
This signature is representative of the randomness encoded in its state due to manufacturing variations and other physical disorders. 
In order to compromise the integrity of the CRP protocol, one has to model this randomness to generate a response representative of the PUF's signature. 
There are two approaches to this problem: a model-based solution and a model-agnostic solution. 
The model-based solution, explored in \cite{Ruhrmair2010}, attempted to capture this randomness through modeling the characteristics of a PUF using domain knowledge (PUF architecture) and characteristics (delay model, thermal response characteristics, etc.). 
Thus, the attack consists of regression of the model's parameters. 

We, however, consider an architecture independent approach to the solution by disregarding the need for a characteristic equation for the PUF. 
We postulate that the challenge and subsequent response of any given PUF is representative of its characteristic function. 
Thus, modeling the dependency between the various features of a given challenge along with the target response allows us to capture the randomness of a given PUF architecture.
To this end, we use several approaches to capture the dependency between the challenge and response pairs of various PUF architectures. 
Since the underlying dependency is not known to be linear or non-linear, we explore several different machine learning models that characterize the dependency with a linear decision boundary (logistic regression) or with a non-linear decision boundary (random forest and artificial neural networks). 

The attack model consists of learning the optimal function that maps the given $n$-bit challenge $C=c_1,c_2,\ldots ,c_n$ to an appropriate output response $R \in \{-1,1\}$ with a probability $p(R|C)$. 
The objective of the attack model is to learn the function $f: C \rightarrow R$ such that the difference between the generated and actual response of the PUF is minimized. Hence the best attack model is characterized by the search for the optimal function $f$ given by
\begin{equation}
%\vspace*{-1ex}
\argmin_{(C_s,R_s)} E[(\hat{f}(C) - f(C))^2]
\label{optimizer_Attack}
\vspace*{-1ex}
\end{equation}
where $\hat{f}(C)$ is the characteristic function of the given PUF architecture and $(C_s, R_s)$ represents the space of all known challenge-response pairs obtained through the eavesdropping protocol. We search for the optimal function $f(C)$ through the characteristic equation of the different machine learning models defined above. For example, in a logistic regression model, $f$ is defined as
\begin{equation}
\vspace{-1ex}
f = \argmax(\sigma(R \times d(\vec{w},C)))
\end{equation}
where $\vec{w}$ is a learned vector that represents the decision boundary ($d$) for the logistic regression model and $\sigma$ is the logistic function.

\subsection{Discriminator Model}
\begin{figure}
\centering
\includegraphics[width=0.99\columnwidth]{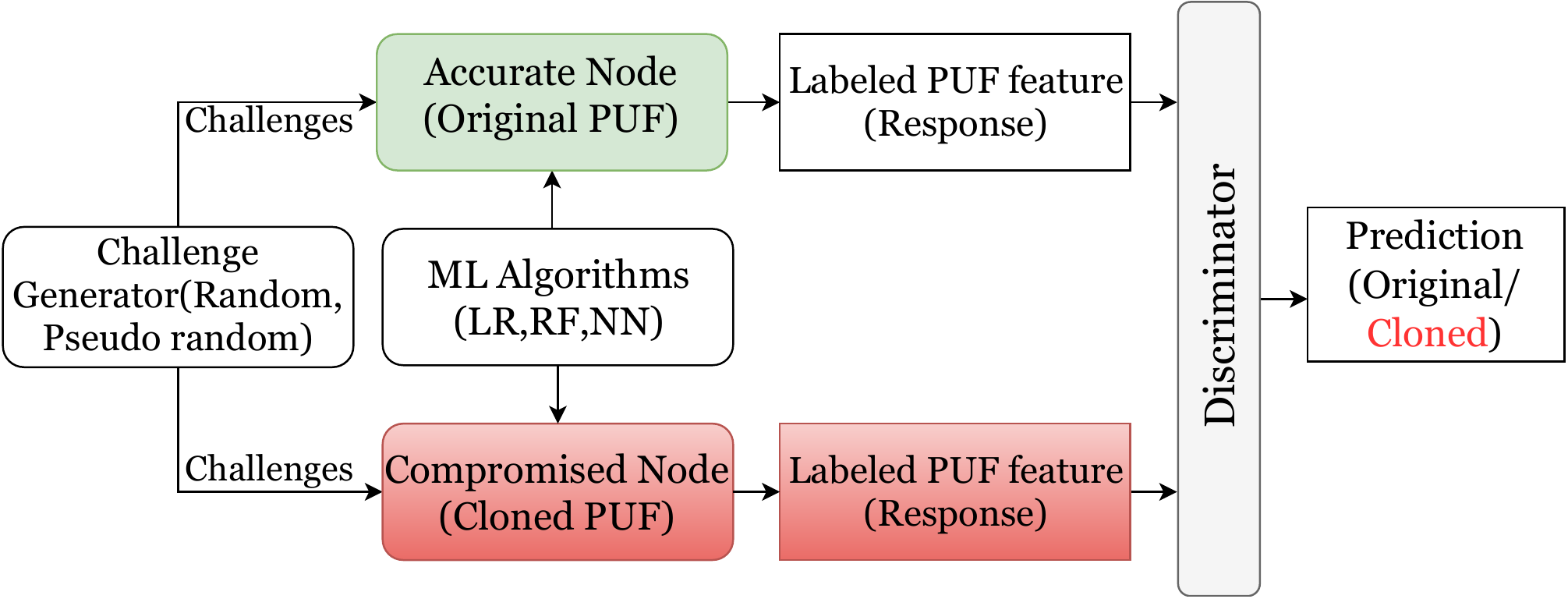}
\caption{ML-based discriminator model to ascertain a PUFs integrity}
\label{fig:defense}
\vspace*{-4ex}
\end{figure}

The modeling of the internal randomness of a given PUF architecture puts the integrity of the CRP-based authentication into question. 
Hence, it becomes critical that we are able to differentiate between the original PUF and an adversarial attack, such as one described in Section \ref{attack}. 
To this end, we introduce a mathematical model that is able to discriminate between an original and a cloned PUF called the \textit{discriminator model}, as illustrated in figure \ref{fig:defense}. 
% The discriminator model is a classifier trained using the supervised learning. 
The discriminator decides whether each instance of the response belongs to the actual PUF or a malicious attacker. 
% The goal of the discriminator, when shown an occurrence from the true response by the original PUF, is to recognize them as original. 
As seen in figure \ref{fig:defense}, the discriminator model takes in the response of the original PUF along with the response of the PUF cloned with several ML attacks as the input to predict whether the PUF is an original or a cloned and returns the probabilities. 
The cloned part of the response is shown in red. 
The output of this discriminator is a single scalar value $D(C)$, indicative of an adversarial attack. 
The value $D(C)$ is a probability function that maps a given response ($R$) to the distribution belonging to either the original PUF ($\hat{f}(C)$) or an attacker ($f(C)$) for a given $n$-bit challenge $C$. Hence, the optimal discriminator model is given by 

\begin{equation}
D^{\star}(C, R) = \frac{p(\hat{f}(C)}{p(f(C)) + p(\hat{f}(C))}
\label{generatorEqn}
\end{equation}

where $D^{\star}(C,R)$ is a mathematical model that maps the response $R$ for a given challenge ($C$) into the probability space of either the original PUF ($\hat{f}$(.)) or the attack model ($f(.)$). 
Again, we explore the use of well-known machine learning models as the basis for our discriminator mathematical model.
% for the linear (logistic regression) and the nonlinear (random forests and neural networks) subspaces.

The search space for the optimal discriminator is similarly characterized by the optimization function defined in Equation \ref{optimizer_Attack}. However, the search is represented by the discriminator to distinguish between the original PUF's response and an adversarial attack.

\subsection{Search for Optimal Attack-Discriminator Model}
The search space for the optimal attack model and discriminator model is defined by the optimizer functions defined in Equation \ref{optimizer_Attack} and its subsequent adaptation for the discriminator, respectively. 
We employ a simple grid search algorithm to find the optimal attack model ($f(.)$) from a given set of possible models ($F$). 
The attack models space, $F$ comprises of all transformation functions that satisfy the condition $f:C \rightarrow R$. 
We restrict the search space to the given three machine learning models: Logistic Regression (LR), Random Forest (RF), and Neural Network (NN). 
We also ensure that the optimal discriminator is chosen from a set of discriminative function $G(.) \in G_s$, where $G_s$ is the collection of all discriminative functions that optimize the probability function defined in Equation \ref{generatorEqn}. 
Again, we restrict the search space to the three aforementioned models. 
While the grid search suffers from the curse of dimensionality and does not scale to large search spaces of $F$ and $G_s$, limiting the number of plausible functions allows us to exhaustively search for the optimal discriminator for a given attack model and a target PUF. 
Additionally, the grid search is a reasonable approach given that it can be embarrassingly parallel.

\begin{figure*}
\centering
\resizebox{\textwidth}{!}{
\begin{tabular}{c c c c}
\includegraphics[width=\textwidth]{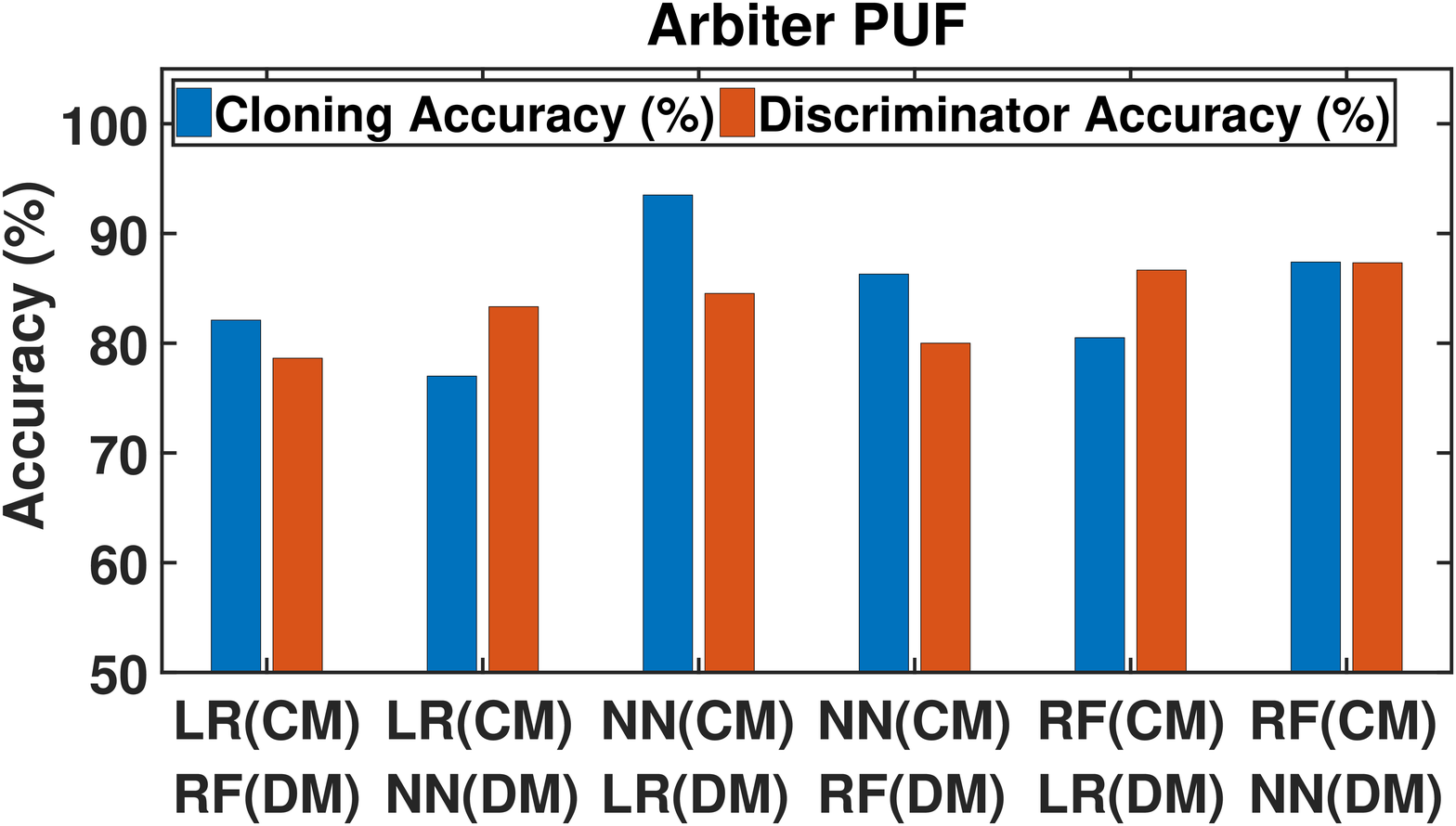} & 
\includegraphics[width=\textwidth]{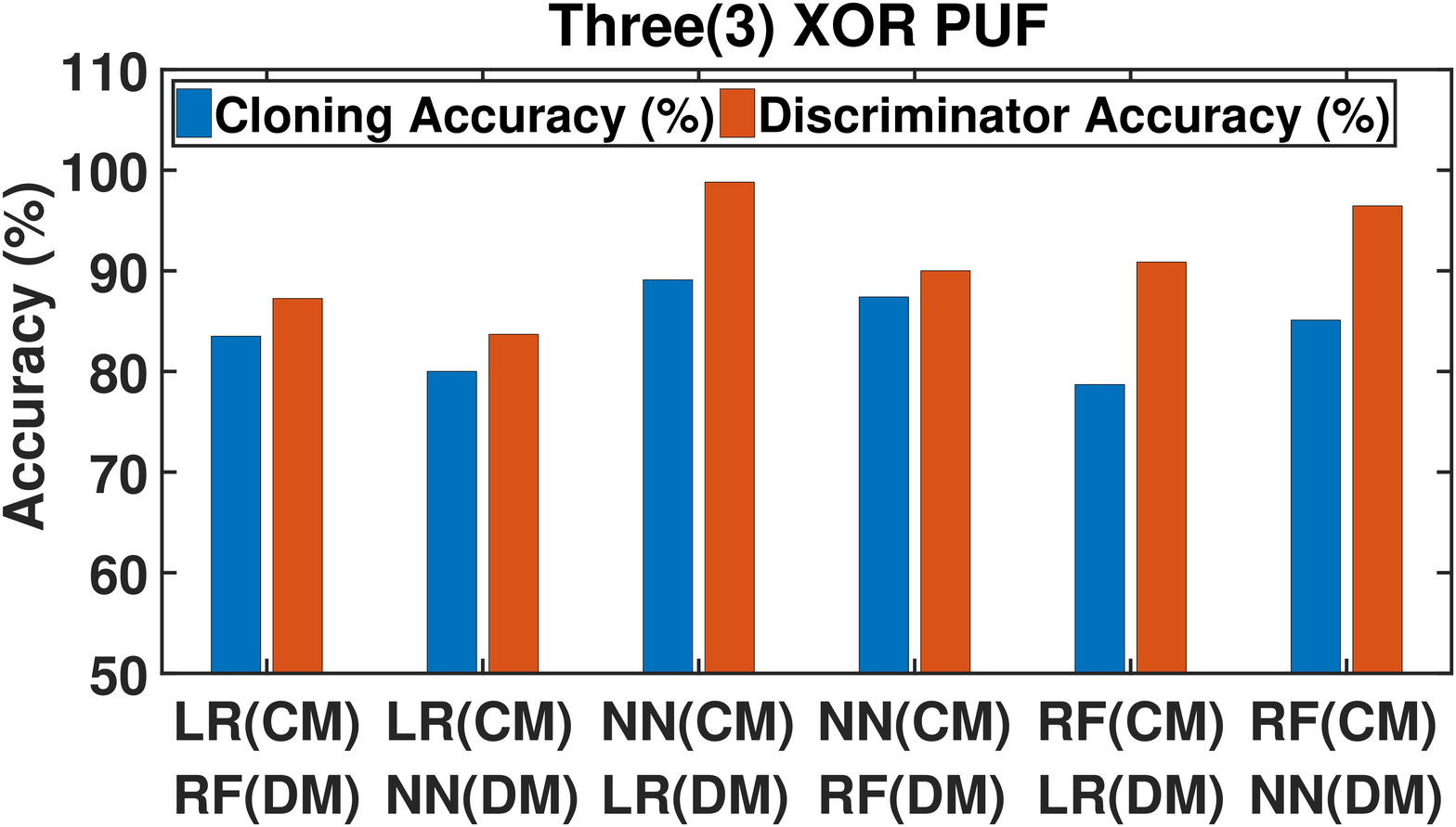} & 
\includegraphics[width=\textwidth]{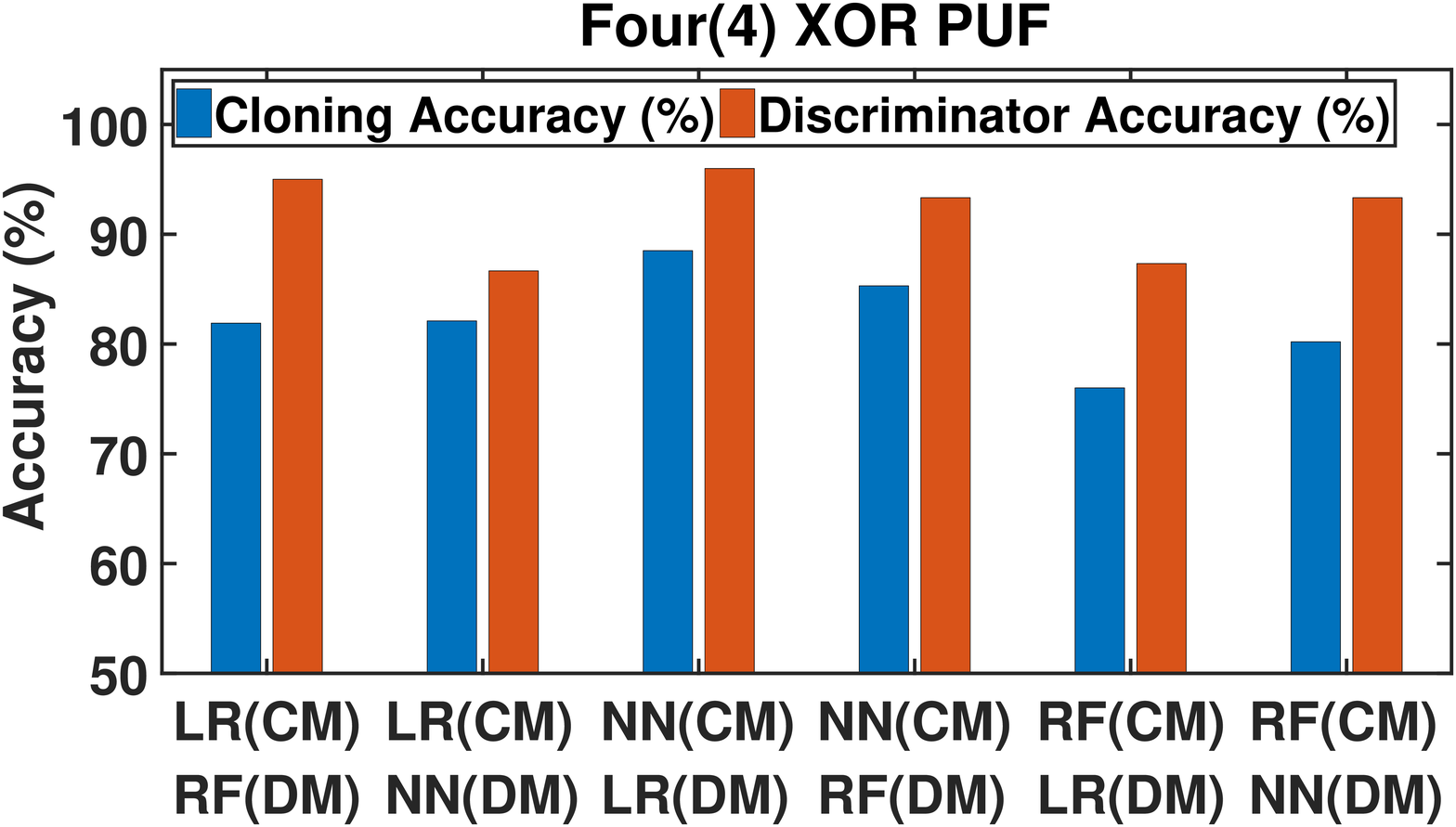} & 
\includegraphics[width=\textwidth]{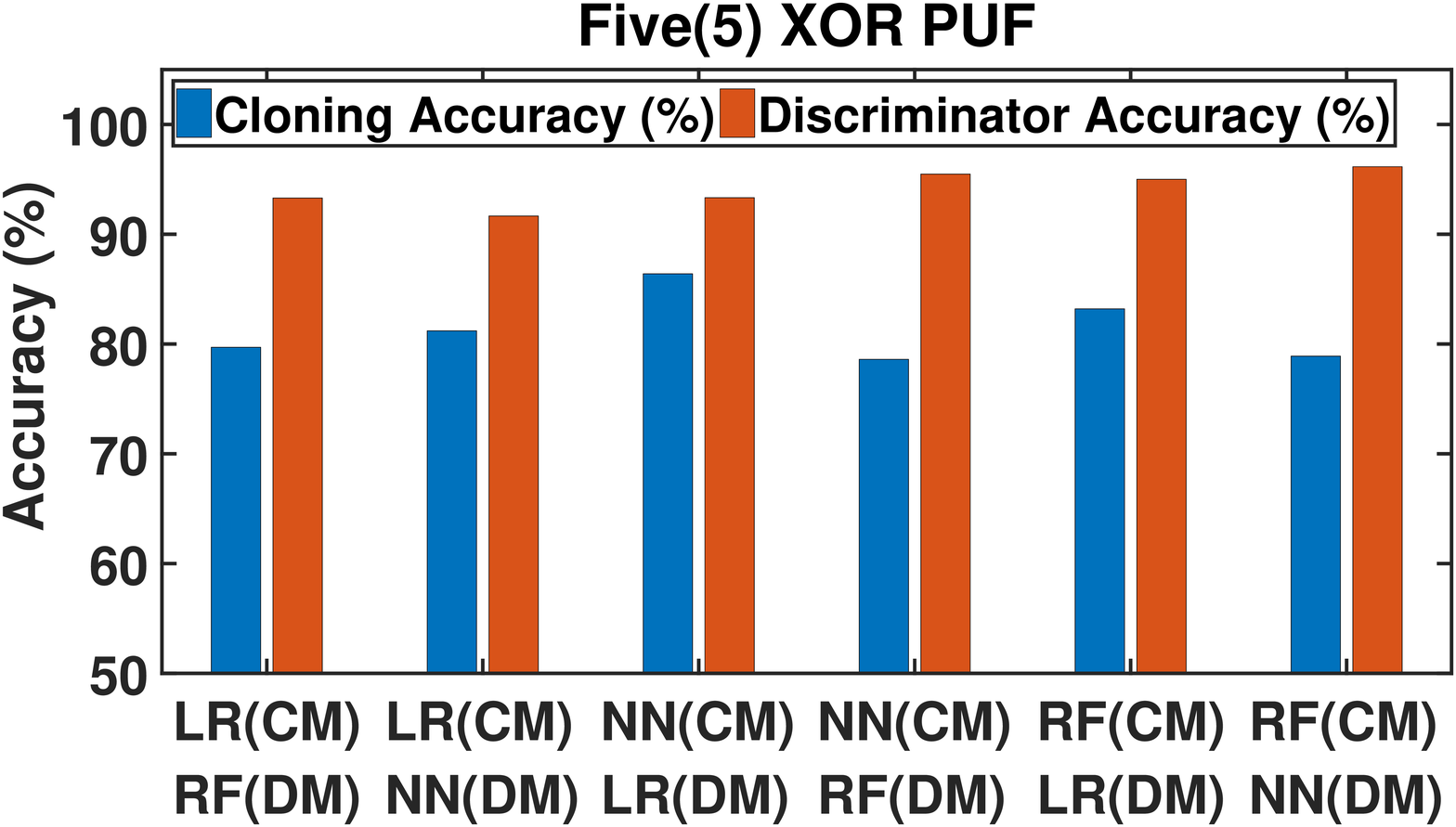} \\
{\Huge (a)} & {\Huge (b)} & {\Huge (c)} & {\Huge (d)} \\
\includegraphics[width=\textwidth]{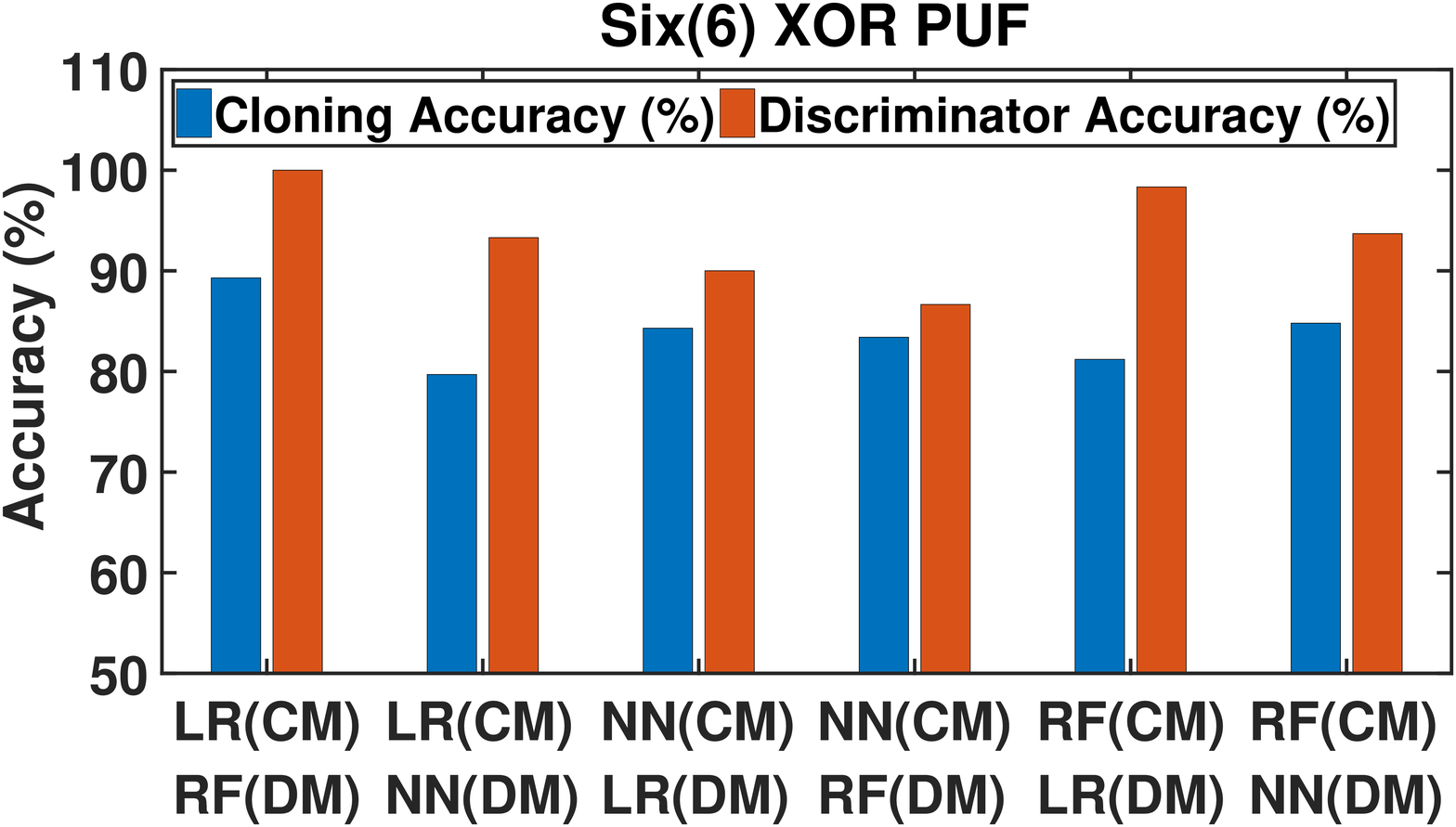} & 
\includegraphics[width=\textwidth]{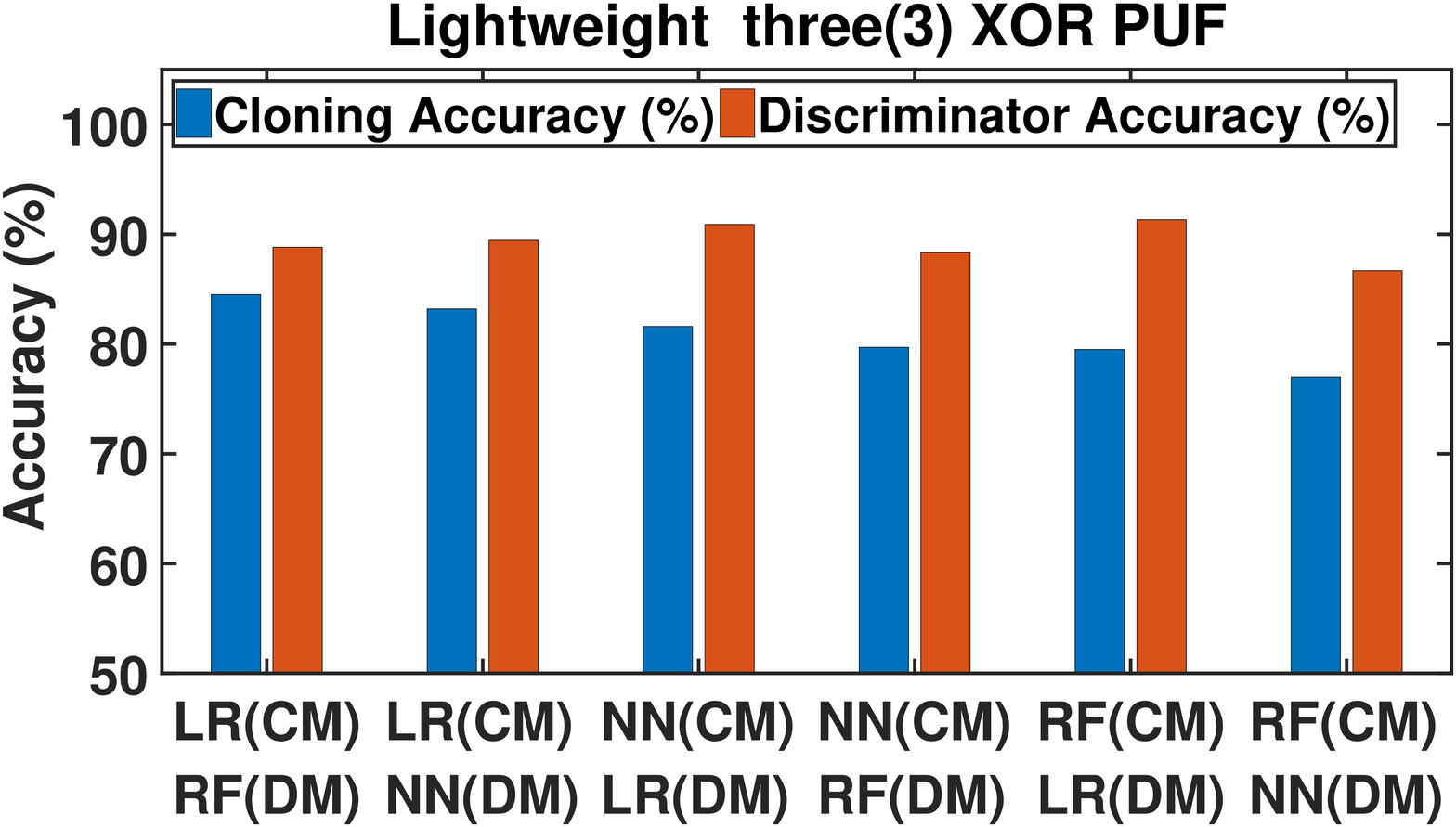} & 
\includegraphics[width=\textwidth]{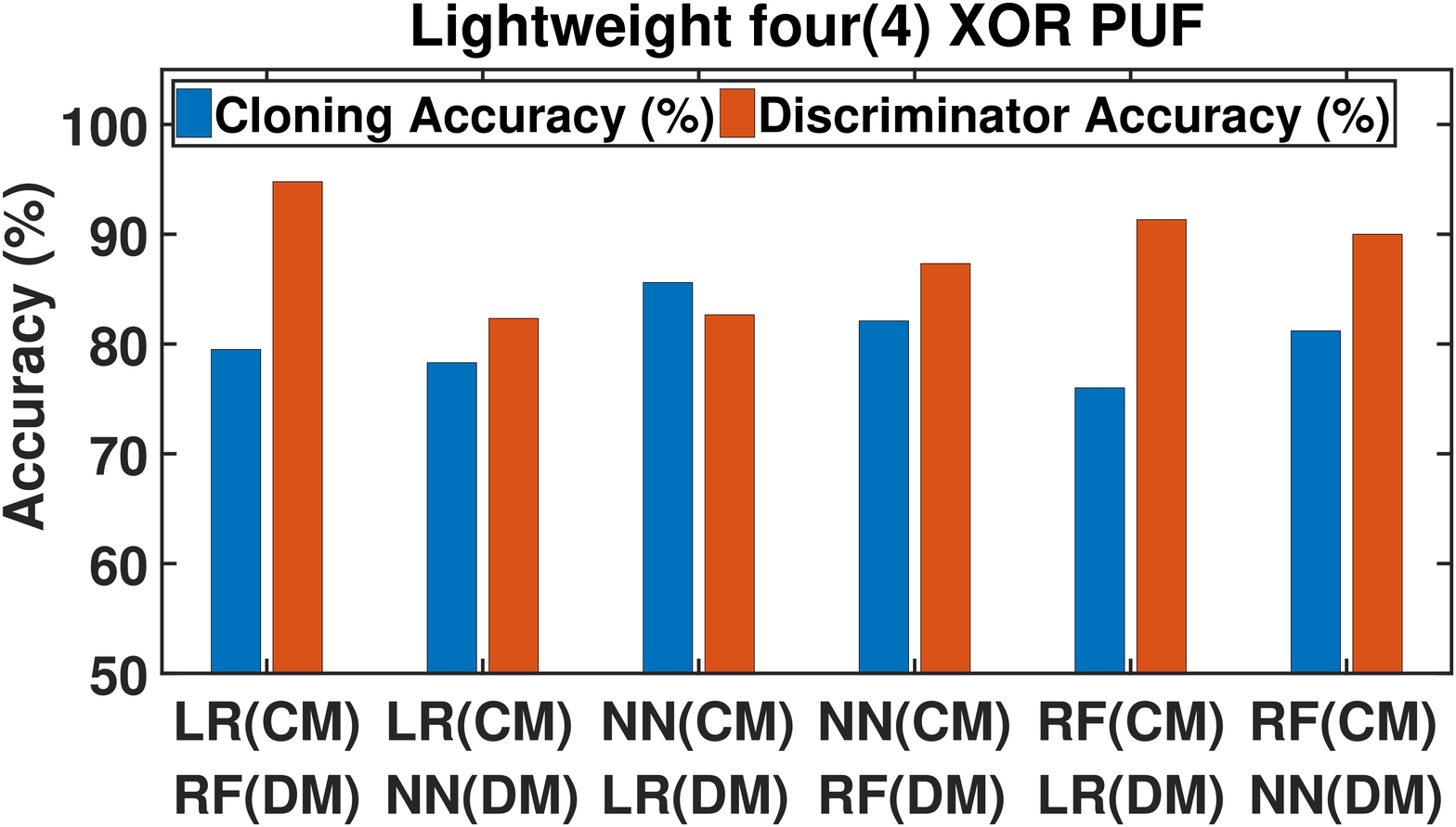} & 
\includegraphics[width=\textwidth]{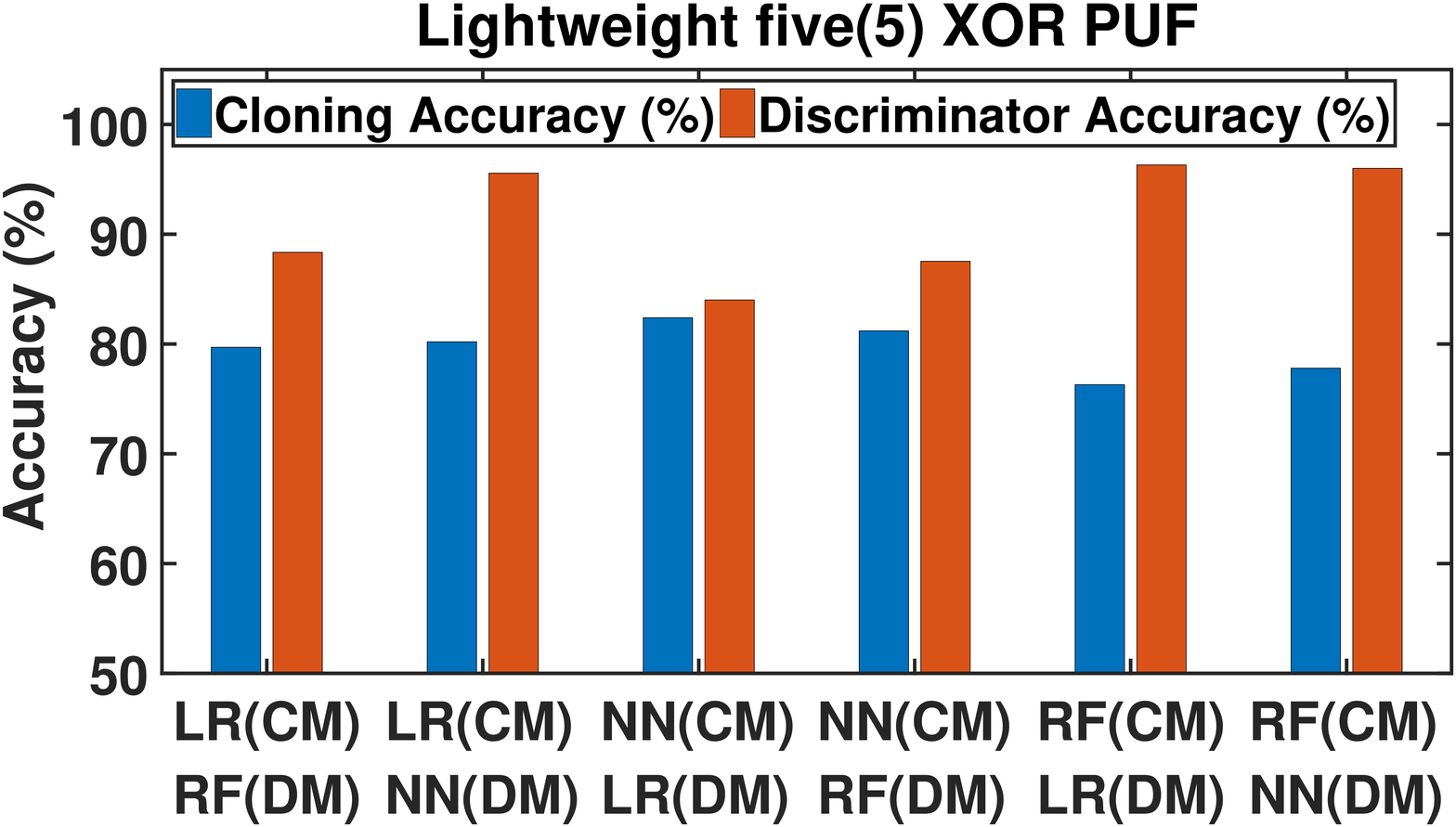} \\
{\Huge (e)} & {\Huge(f)} & {\Huge(g) }& {\Huge(h)} \\
\end{tabular}}

\caption{ Comparison of cloning and discriminator accuracy for different PUFs architecture with combinations of ML models. Single bar represents average accuracy for 64, 128, and 256 stages.  Along X-axis, X(Y) defines X model is used for Y task where Y can be cloning (CM) or discriminator (DM). }
\label{fig:comboplot}
% \vspace{-1ex}
\end{figure*}

\begin{figure*}
\centering
\resizebox{\textwidth}{!}{
\begin{tabular}{c c c c}
\includegraphics[width=\textwidth]{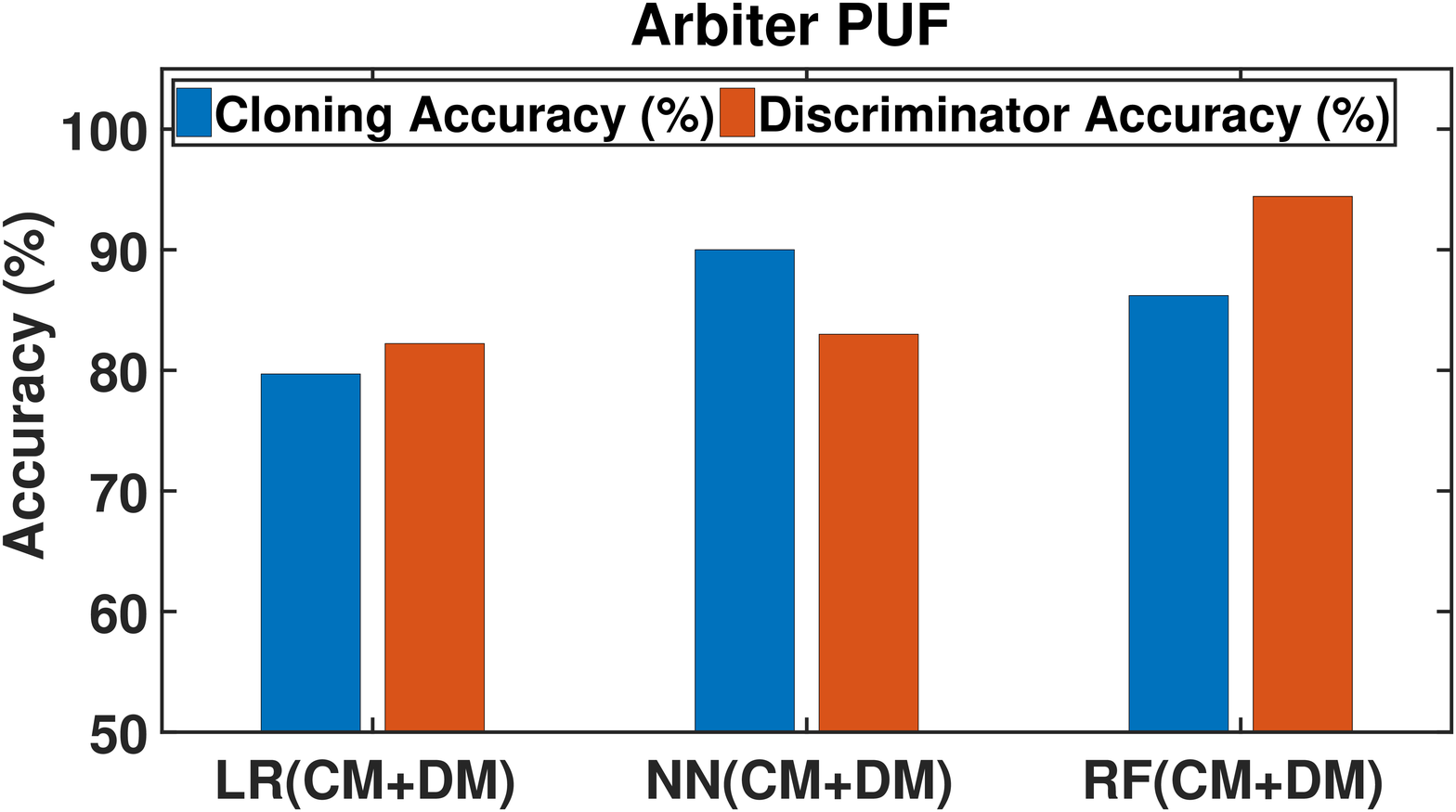} & 
\includegraphics[width=\textwidth]{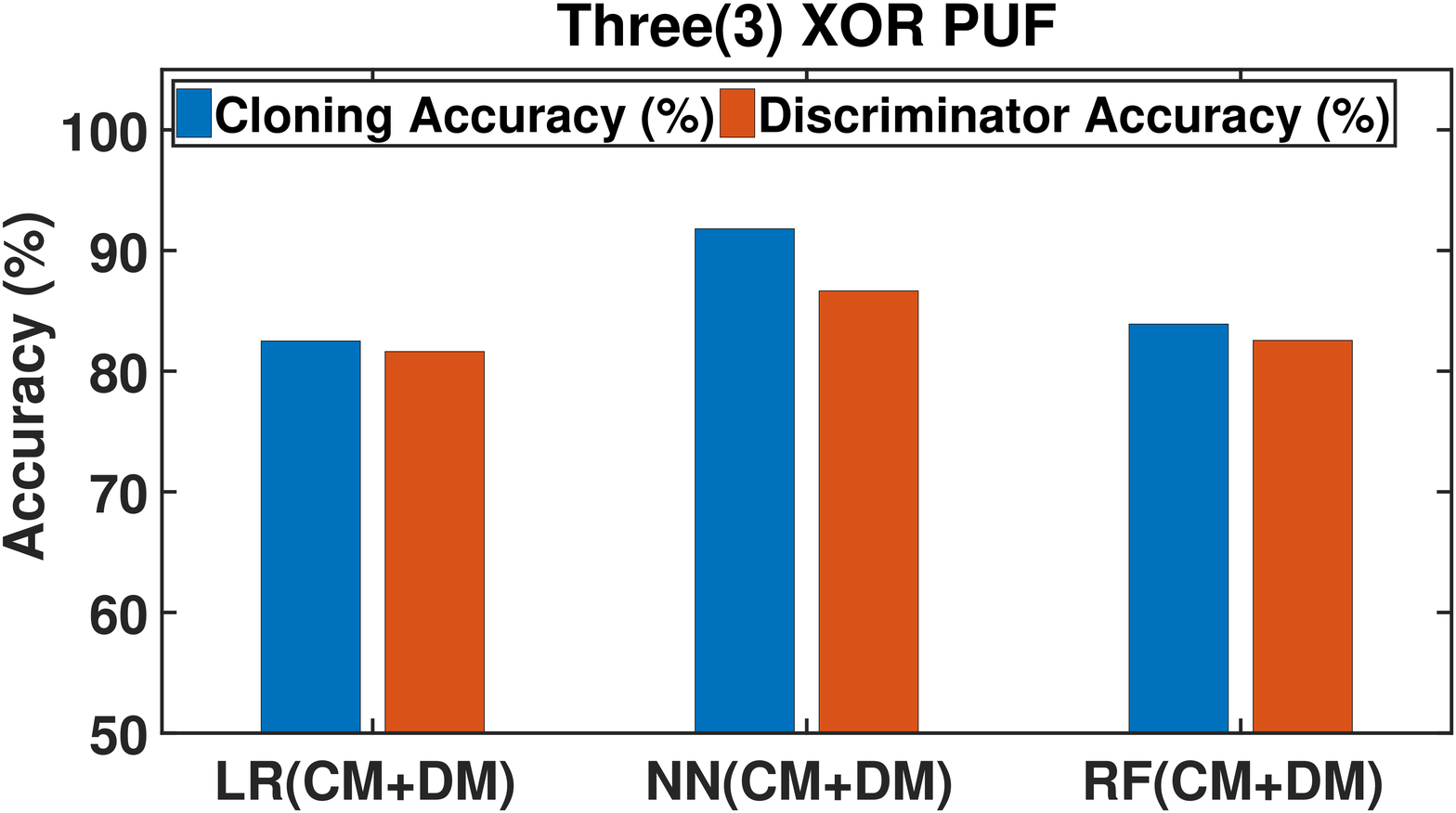} & 
\includegraphics[width=\textwidth]{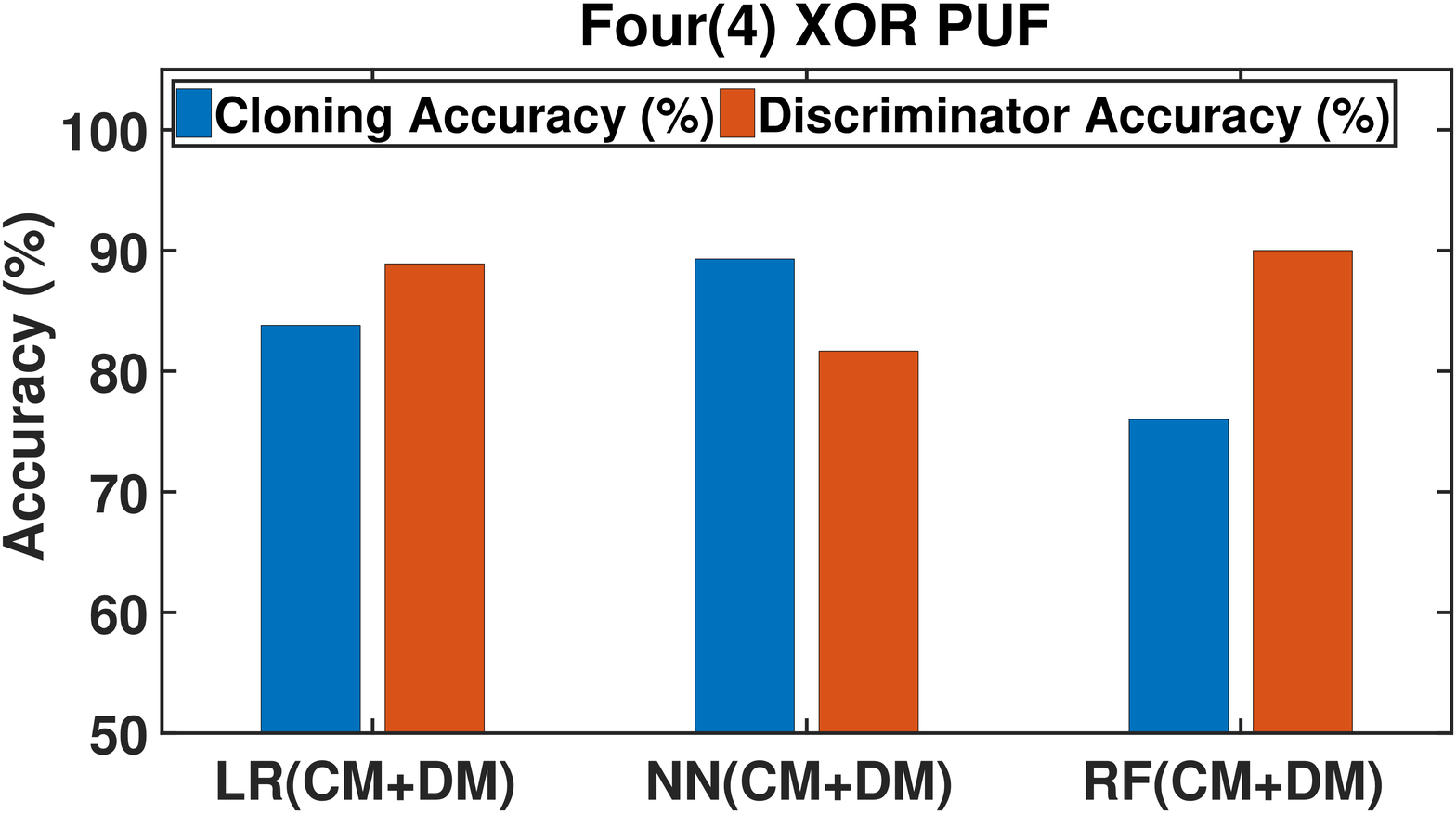} & 
\includegraphics[width=\textwidth]{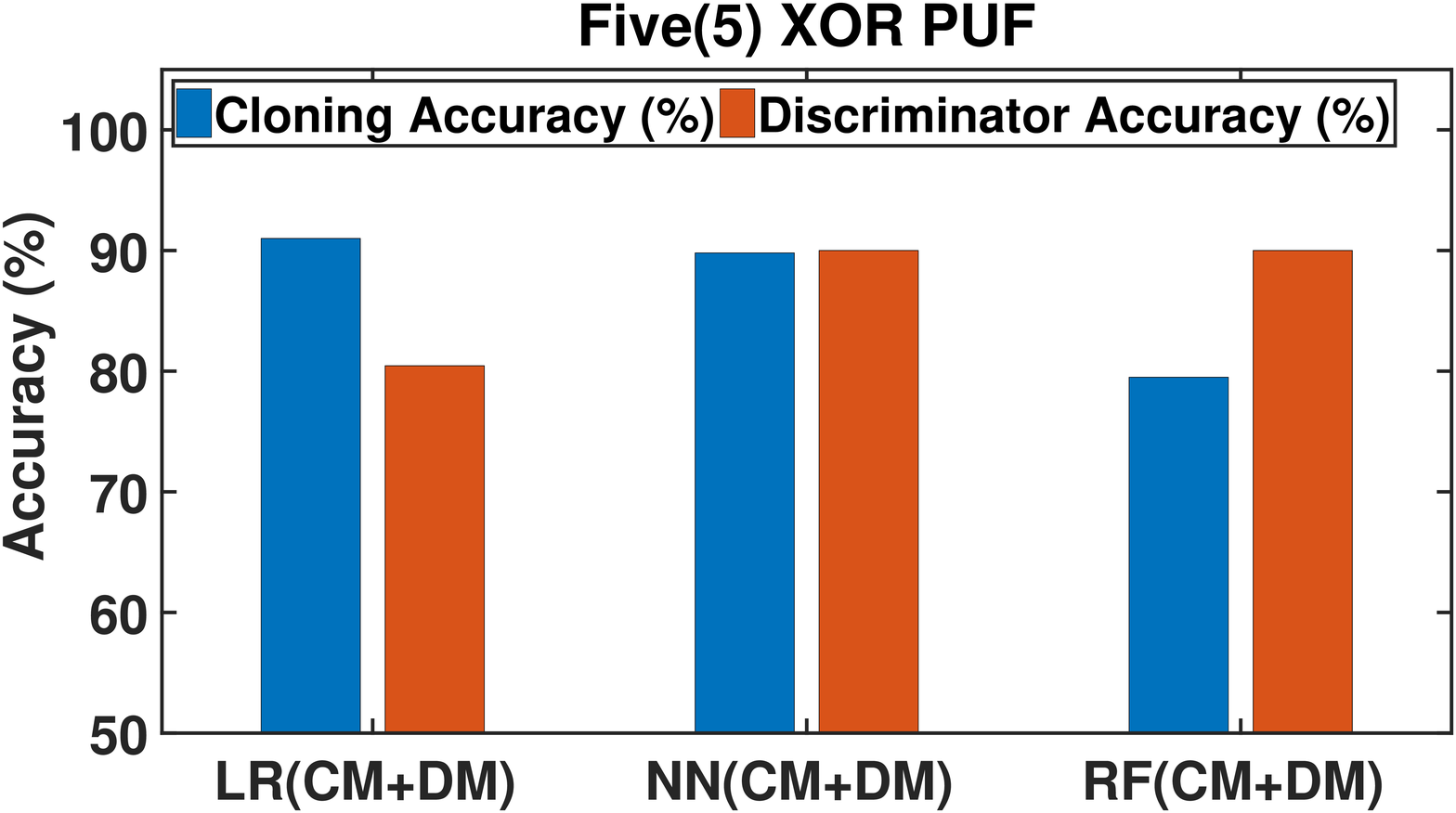} \\
{\Huge (a)} & {\Huge (b)} & {\Huge (c)} & {\Huge (d)} \\
\includegraphics[width=\textwidth]{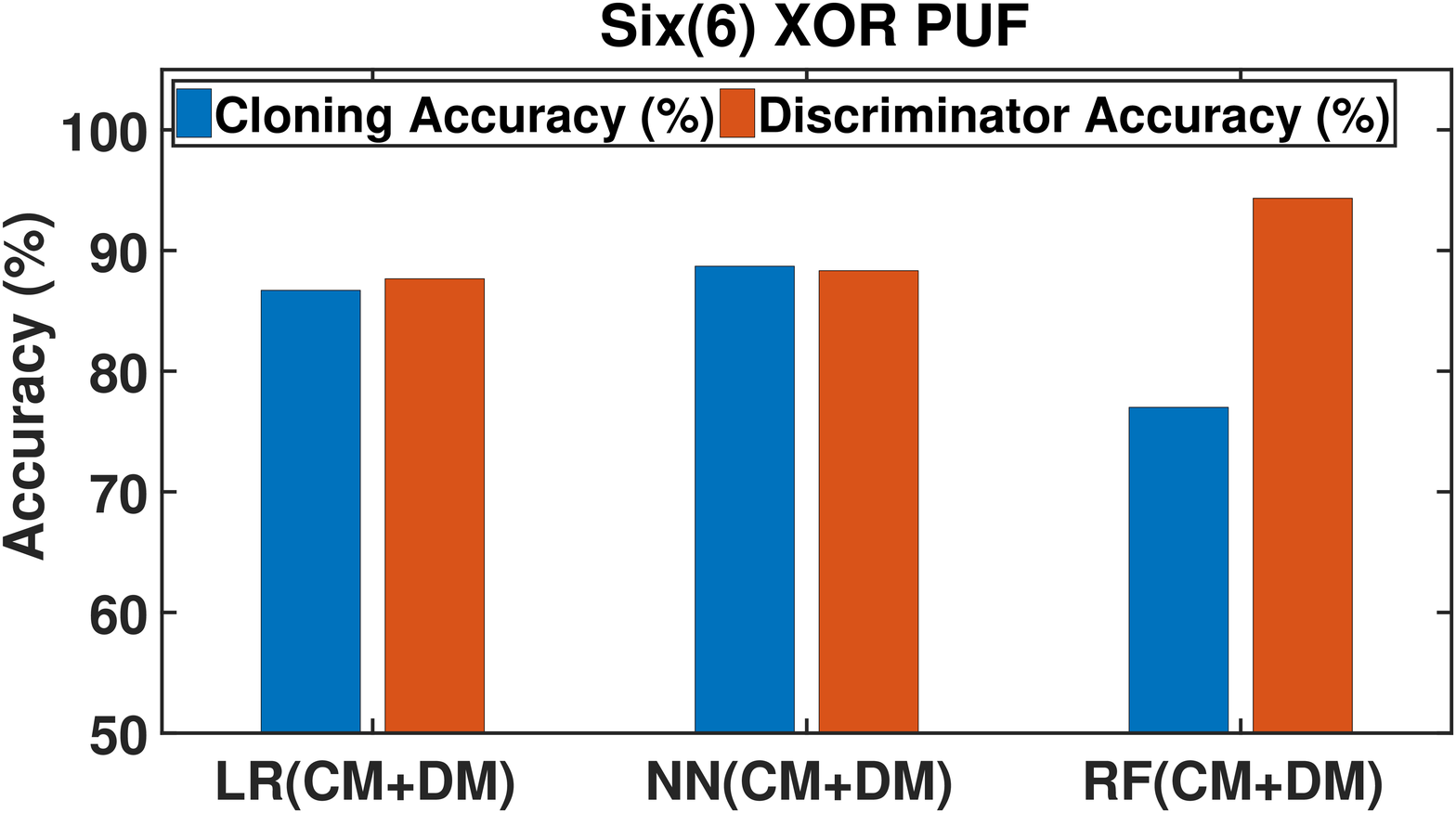} & 
\includegraphics[width=\textwidth]{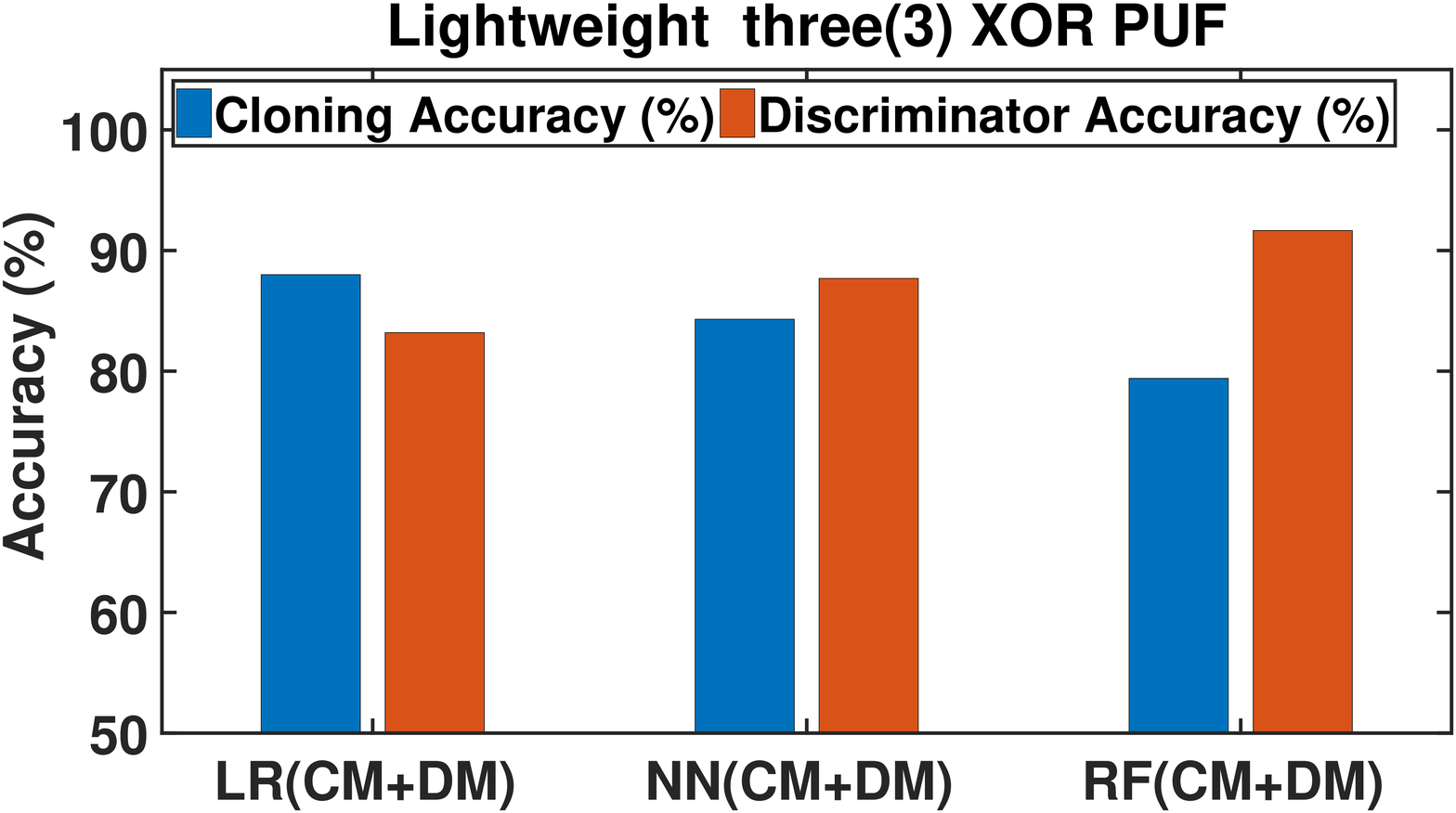} & 
\includegraphics[width=\textwidth]{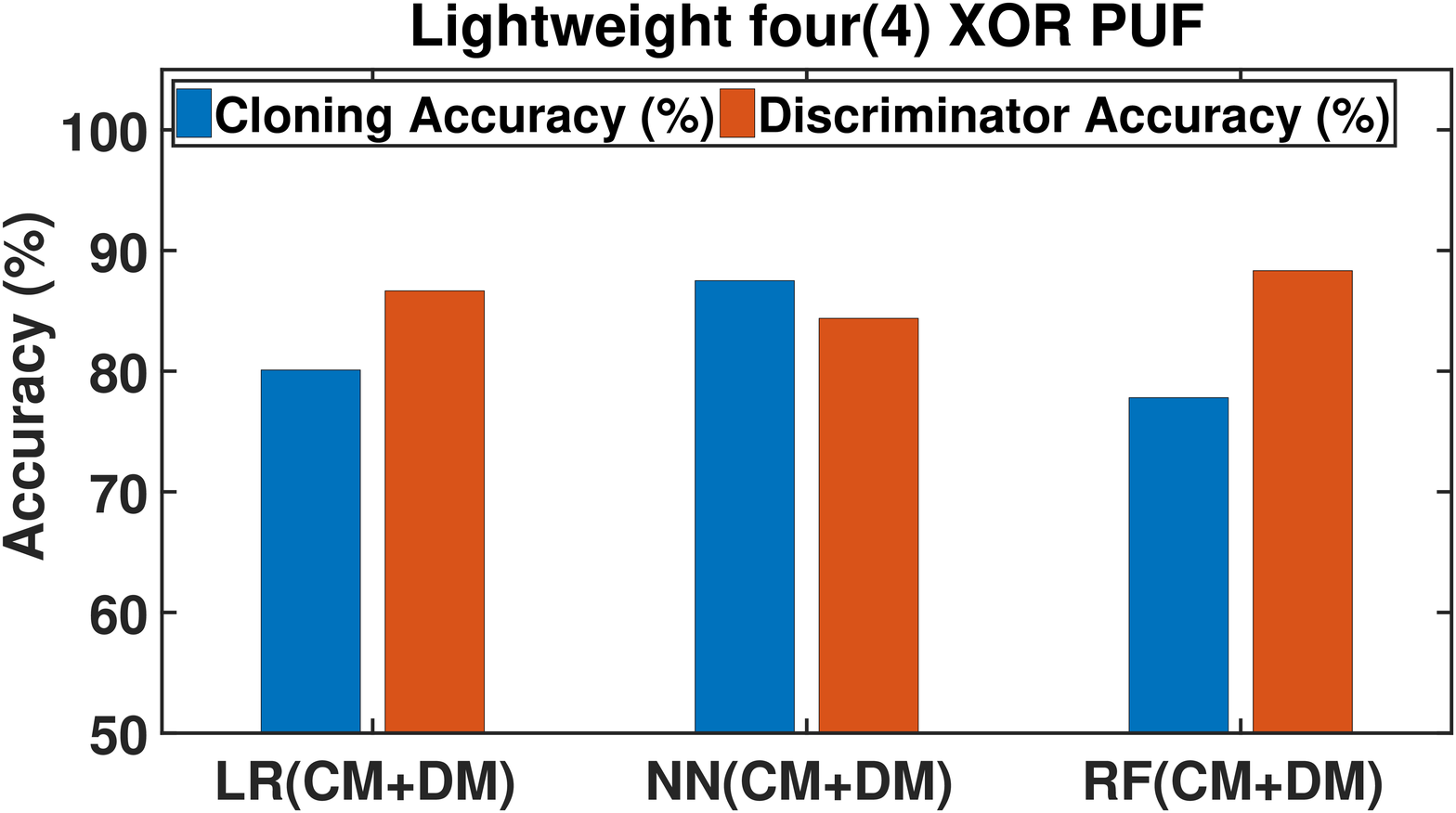} & 
\includegraphics[width=\textwidth]{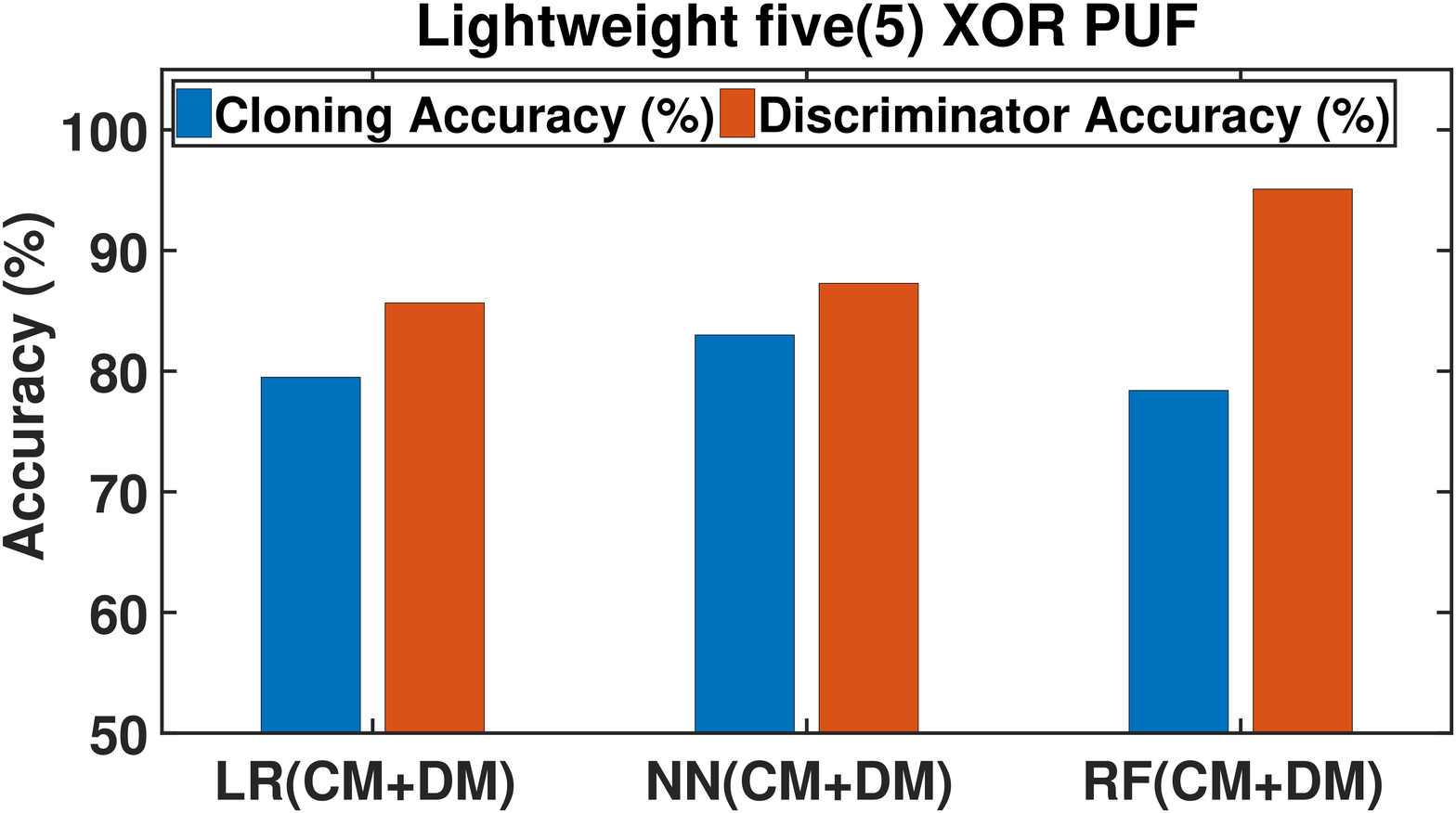} \\
{\Huge (e)} & {\Huge (f)} & {\Huge (g)} & {\Huge (h)} \\
\end{tabular}}

\caption{Comparison of cloning and discriminator accuracy for different PUFs architecture under single ML model. Single bar represents average accuracy for 64, 128, and 256 stages.  Along
X-axis, X(Y) defines only X model is used for Y where Y defines both cloning (CM) and discriminator (DM) tasks.}
\label{fig:singleplot}
\vspace{-3ex}
\end{figure*}

%%%%%%%%%%%% Error estimation %%%%%%%%%%%%%%%%%%%%%%%%%
\begin{table}
\begin{center}
\caption{ML Model error estimation for cloning and discriminator and cloning time}
\label{tab:error}
\vspace*{-4ex}
\resizebox{\columnwidth}{!}{
\begin{tabular}{l|c|c|c}
\toprule
PUF Model  &   {Cloning Error (\%)} &   {Discriminator Error (\%)} & {Cloning Time}\\
\midrule\midrule
APUF &  6.50\% & 12.66\% & 0.002 sec \\ 
3 XOR APUF & 8.20\% & 1.18\% & 70:85 sec  \\
4 XOR APUF &  10.70\% & 4.03\% & 1:38 min\\     
5 XOR APUF & 9.00\% & 3.84\% & 62:48 min\\
6 XOR APUF* & 10.70\% & 0.50\% & 240 min\\
LW 3 XOR APUF & 12.00\% & 8.66\% & 1:59 sec\\
LW 4 XOR APUF & 12.50\% & 5.22\% & 30:58 min\\
LW 5 XOR APUF* & 17.00\% & 3.69\% & 180 min\\
\hline
Average & 10.83\% & 4.97\%\\
\bottomrule
\end{tabular}}
\end{center}
*Note that in the literature \cite{Ruhrmair2010}\cite{Ruhrmairmulti}, the maximum number of
XORs used is 6.  It is known that 6 XORs is sufficient
to give a strong PUF.
\vspace{-4ex}
\end{table}

\section{Experimental Results and Discussion}
\label{sec:results}
\vspace{-0.4ex}
% We did not find universally accepted CRPs for oblivious PUF models as a way to attack. 
Following experimental setup by \cite{Ruhrmair2010}, we report the upper bound of attacker and discriminator accuracy in a supervised setting. We consider three strong PUF architectures (Arbiter, XOR and Lightweight), while each of them contains three stages (64, 128, and 256) and the number of XOR is limited to (3, 4, and 5) for both XOR and Lightweight PUFs. 
This gives us a total of 24 different strong PUF architectures for  validating the efficacy of the proposed method. 

The average cloning error, the discriminator error and the cloning time is shown in Table \ref{tab:error}. We report average results over 5 different runs for each PUF architecture. 
From Table \ref{tab:error}, we can see that on average, a strong PUF can be cloned with a cloning error of 10.83\% irrespective of its underlying architecture of the PUF. 
The discriminator error shows that the obtained response is either from the original PUF or from the cloned PUF with an error of 4.97\%. %The discriminator error indicates the ability of the model to differentiate a compromised node and the original node. 
The aging of the PUF \cite{meguerdichian2011device} affects the delay characteristic which produces a different pattern of the responses compared to the compromised node. It can seen that the cloning time is reasonable, particularly given the complexity and stochastic nature of the considered PUFs.
We discuss the performance of our approach on different PUF architectures in detail below.

% 4.a. What happened in their experiments?
% 4.b. What are the major results of their experiments
% that are related to the impact you are studying?

% The results from both combined and individual ML algorithms performance on attacker and discriminator prediction accuracy are shown in Figs. \ref{fig:comboplot} and \ref{fig:singleplot}.

\textbf{Arbiter PUF:}
As seen from Figures \ref{fig:comboplot}(a) and \ref{fig:singleplot}(a), modeling attack on the exact number of stages for arbiter PUF can be accurately guessed (93.5\%) for a combination of NN(CM) and LR(DM). 
% Remember that, our underlying PUFs architecture contains 24 different models for PUF parameters extraction. 
However, with a single ML model (RF), the discriminator prediction accuracy improves to (94.4\%) compared with combined NN(CM) and LR(DM). 

\textbf{XOR PUF:}
Figures \ref{fig:comboplot}(b) and \ref{fig:singleplot}(b) show the ML models performance for three (3) XOR PUF. While the combined NN(CM) and NN(DM) models are capable of capturing the PUF parameters 91.8\% of the time, we are able to make use of combined NN(CM) and LR(DM) to get discriminator accuracy up to 98.8\%.
It can be seen from Figures \ref{fig:comboplot}(c) and \ref{fig:singleplot}(c) that the single NN model performance is markedly higher (89.3\%) in modeling attack compared to all other combinations for four (4) XOR PUF. Similarly, the combined NN(CM) and LR(DM) outperforms others in discriminator prediction with 95.9\% accuracy. 
As shown in Figures \ref{fig:comboplot}(d) and \ref{fig:singleplot}(d), our trained single ML model (LR) notably improves modeling attack accuracy about 91.0\% for five (5) XOR PUF. The discriminator accuracy improves up to 96.2\% for combined RF(CM) and NN(DM).
Finally, for six (6) XOR PUF in Figures \ref{fig:comboplot}(e) and \ref{fig:singleplot}(e), the combined LR(CM) and RF(DM) can achieve up to 89.3\% and 100\% for modeling the attack and discriminator, respectively.

\textbf{Lightweight PUF (LPUF):}
For LPUF, we evaluate our method on 3-, 4-, and 5-XOR PUF. With supervised experiment for LPUF with three (3) XOR, the single LR model can achieve modeling accuracy up to 88.0\% while that for discriminator, the single RF outperforms all other combinations by 91.7\% as shown in Figures \ref{fig:comboplot}(f) and \ref{fig:singleplot}(f).
Figures \ref{fig:comboplot}(g) and \ref{fig:singleplot}(g) show modeling and prediction accuracy for LPUF with four (4) XOR. In this case, the single NN improves cloning accuracy by 87.5\% and the combined LR(CM) and RF(DM) improves the discriminator performance by 94.8\%.
Finally, we apply the proposed method for LPUF with five (5) XOR as shown in Figures \ref{fig:comboplot}(h) and \ref{fig:singleplot}(h). For a single NN model, the cloning accuracy is 83.0\% and the discriminator prediction can improve by 96.3\% the combined RF(CM) and LR(DM). 
\section{Conclusion}
\label{sec:conclusions}

In this work, we introduced an efficient architecture-independent machine learning based approach for cloning strong PUFs. We also introduce a novel discriminator model to identify cloned and original PUFs with a high degree of confidence. We also introduce a search-based approach for identifying the optimal discriminator model for a given cloned PUF using three common ML models. Extensive experiments show the efficacy of the proposed approach. For future work, we will extend this method for control PUFs and explore ensemble meta-algorithms.  

% We perform a  three (3) ML models to evaluate these tasks. We have also demonstrated that a combination of ML models can sometimes outperform single ML model and vice-versa. The competitiveness of the proposed technique can be further improved as a better classifier with more labeled training data. 
% For future work, we plan to validate the proposed work with emulated PUF models on real hardware  as well as ensemble meta-algorithms, for better attack and defense mechanism for  IoT edge devices using PUF for authentication.

% 5.1. What are the implications of the work?
% 5.2. Why do you believe this paper has impact?
%such as ring oscillator PUFs

\bibliographystyle{unsrt}
\scriptsize{
\bibliography{bib/model_attack,bib/puf_basic,bib/model_prevent,bib/miscell,bib/IoT}
}

\end{document}